\documentclass[iop]{emulateapj}
\usepackage{multirow}
\usepackage{bm}
\usepackage{amssymb}
\usepackage{amsmath}
\usepackage{latexsym}
\usepackage{graphicx}
\usepackage{ifsym}
\usepackage{bm}
\usepackage{tabularx}
\usepackage{hyperref}
\usepackage{enumitem}

\slugcomment{}
\shorttitle{ALMA and MagAO Observations of GQ Lup B}
\shortauthors{Wu et al.}

\begin{document}
\title{An ALMA and M\lowercase{ag}AO Study of the Substellar Companion GQ Lup B\footnotemark[$\ast$]}
\footnotetext[$\ast$]{This paper includes data gathered with the 6.5 m Magellan Clay Telescope at Las Campanas Observatory, Chile.}

\author{Ya-Lin Wu$^1$, 
 Patrick D. Sheehan$^1$,
 Jared R. Males$^{1}$,
 Laird M. Close$^1$, 
 Katie M. Morzinski$^{1}$, 
 Johanna K. Teske$^{2,}$\footnotemark[$\dagger$],\footnotetext[$\dagger$]{Carnegie Origins Fellow, joint appointment between Carnegie DTM and Carnegie Observatories.}
 Asher Haug-Baltzell$^{3}$,
 Nirav Merchant$^{3,4}$, and
 Eric Lyons$^{3,4}$
}

\affil{$^1$Steward Observatory, University of Arizona, Tucson, AZ 85721, USA; yalinwu@email.arizona.edu\\
$^2$Department of Terrestrial Magnetism, Carnegie Institute of Washington, 5241 Broad Branch Road, NW, Washington, DC 20015, USA\\
$^3$CyVerse, University of Arizona, Tucson, AZ 85721, USA\\
$^4$Bio5 Institute, University of Arizona, Tucson, AZ 85721, USA\\
{\it Accepted for publication in ApJ}}

\begin{abstract} 
Multi-wavelength observations provide a complementary view of the formation of young directly-imaged planet-mass companions. We report the ALMA 1.3 mm and Magellan adaptive optics (MagAO) H$\alpha$, $i'$, $z'$, and $Y_S$ observations of the GQ Lup system, a classical T Tauri star with a 10--40 $M_\text{Jup}$ substellar companion at $\sim$110 AU projected separation. We estimate the accretion rates for both components from the observed H$\alpha$ fluxes. In our $\sim$$0\farcs05$ resolution ALMA map, we resolve GQ Lup A's disk in dust continuum, but no signal is found from the companion. The disk is compact, with a radius of $\sim$22 AU, a dust mass of $\sim$6 $M_\earth$, an inclination angle of $\sim$56\degr, and a very flat surface density profile indicative of a radial variation in dust grain sizes. No gaps or inner cavity are found in the disk, so there is unlikely a massive inner companion to scatter GQ Lup B outward. Thus, GQ Lup B might have formed in situ via disk fragmentation or prestellar core collapse. We also show that GQ Lup A's disk is misaligned with its spin axis, and possibly with GQ Lup B's orbit. Our analysis on the tidal truncation radius of GQ Lup A's disk suggests that GQ Lup B's orbit might have a low eccentricity. 
\end{abstract}


\section{INTRODUCTION}
In recent years, high-contrast imaging surveys have discovered many wide-orbit substellar companions, which are located at tens to hundreds of AU from their host stars and have masses of a few to tens of $M_\mathrm{Jup}$. Some of these companions have features indicative of accretion disks, including optical and near-infrared emissions such as H$\alpha$, Br-$\gamma$, and Pa-$\beta$ (e.g., \citealt{WG01}; \citealt{S07}; \citealt{S08}; \citealt{Bowler11}, \citeyear{B14}; \citealt{P12}; \citealt{Z14}; \citealt{L15}; \citealt{S15}), high dust extinction (e.g., \citealt{S08}; \citealt{W15a}, \citeyear{W15b}), and infrared excess from dust emission (e.g., \citealt{Bowler11}; \citealt{Ba13}; \citealt{K14}; \citealt{Cu15}). It is expected that disks could be common among young substellar companions because the main formation mechanisms---collapse of prestellar cores, fragmentation of circumstellar disks, and core accretion plus subsequent scattering---can all produce disk-bearing companions. Each mechanism, however, can leave distinct imprints on disk properties. For instance, objects formed by disk fragmentation may have higher disk masses and accretion rates compared to those formed by prestellar core collapse \citep{SH15}. On the other hand, scattering can be destructive to disks (e.g., \citealt{RC01}), and unlike stars, low-mass objects are not efficient to accrete new disks from the natal molecular clouds after scattering \citep{B12}. Characterizing disks of substellar companions therefore provides a new avenue for studying wide companions' mass assembly history.

In addition, if gas emission lines such as CO can be spatially and spectrally resolved, the dynamical mass of the central object can be determined assuming a Keplerian velocity field. Since masses are usually derived by comparing observables to theoretical predictions, this dynamical approach has great potential to calibrate evolutionary models (e.g., \citealt{Cz15,Cz16,M16}). Finally, disk masses, sizes, structure, and lifetimes ultimately regulate satellite formation and satellite-disk interaction. As wide companions are well separated from their host stars, they offer a clear view to the relevant physical processes.

Still, imaging disks around very low-mass companions remains challenging (e.g., \citealt{I14}). Simulations have shown that they tend to be compact because they are tidally truncated at $\sim$1/3 of the Hill radius (e.g., \citealt{QT98,AB09,SB13}). For objects in nearby star-forming regions ($\sim$100 to 150 pc), their disks are probably not larger than $\sim$5 to 30 AU in radii, which in turn requires a $<$0\farcs1 resolution to resolve them in (sub)millimeter. With the advent of ALMA, it is now possible to directly image and characterize these disks. \cite{B15} showed that GSC 6214-210 B has a dust mass of $<$0.15~$M_\earth$ in its disk. \cite{M16} also found that only $<$0.04 $M_\earth$ of dust is present in the disk around GQ Lup B. Most notably, \cite{K15} and \cite{C15} detected FW Tau C's accretion disk in 1.3 mm dust continuum and ${}^{12}$CO (2--1) emission, respectively. \cite{K15} inferred a dust mass of 1--2 $M_\earth$, sufficient to form satellites analogous to the Galilean moons. 

Here we present the ALMA 1.3 mm map of the GQ Lup system, a pre-main-sequence star with a 10--40~$M_{\mathrm{Jup}}$~companion at $\sim$110 AU projected separation. Both components have been shown to exhibit accretion signatures (e.g., \citealt{B01}; \citealt{Z14}). With a $\sim$$0\farcs05$ resolution, we resolve the primary star's accretion disk. The companion's disk is, however, not detected. We also present the 0.6--1~\micron~imaging of the system using the Magellan adaptive optics (MagAO; \citealt{C12}; \citealt{K13}; \citealt{M14}), and derive mass accretion rates from H$\alpha$ intensities.

\section{GQ Lup A and B}
GQ Lup A is a classical T Tauri star (CTTS) with a spectral type of K7eV (\citealt{H62}, \citeyear{H77}) in the Lupus I cloud ($\sim$150 pc; \citealt{C00}; \citealt{F02}). Spectropolarimetric observations suggested that it is 2--5 Myr old, with a photospheric temperature of 4300 K and a mass of 1.05 $M_\sun$ \citep{D12}. Photometric monitoring indicated that it has an inclination of 27$\degr$ and a rotation period of 8.45 days \citep{B07}. It is one of the most studied CTTSs due to many features indicative of active accretion onto the star from a circumstellar disk, such as brightness variations (e.g., \citealt{FS71}; \citealt{A78}; \citealt{B07}; \citealt{D12}), inverse P-Cygni profiles (e.g., \citealt{A78}; \citealt{B82}; \citealt{B01}), optical veiling (e.g., \citealt{B01}; \citealt{SD08}), very intense magnetic field (e.g., \citealt{D12}; \citealt{J13}), and strong excess from near to far-infrared (e.g., \citealt{M68}; \citealt{H94}; \citealt{KS06}; \citealt{M12}; \citealt{MC12}). The accretion disk was first imaged at 1.3 mm dust continuum emission by \cite{D10} using the Submillimeter Array. The disk is compact, low-mass, and possibly tidally truncated. \cite{MC12} presented far-infrared spectra taken with $Herschel$ PACS and proposed that the 63~\micron~excess may come from crystalline water ice in the outer disk. Recently, \cite{M16} presented the ALMA 870~\micron~and CO (3--2) imaging of the disk and derived a gas-to-dust ratio well below typical ratios in the interstellar medium (ISM).

The substellar companion GQ Lup B was discovered by \cite{N05}, and its common proper motion was soon confirmed \citep{MN05,J06,N08}. Mass estimates are very model-dependent, but most studies overlap in the range of $\sim$10 to 40 $M_\mathrm{Jup}$ (see Table 1 of \citealt{L09} for published results). Like the primary star, GQ Lup B is believed to harbor an accretion disk. Lines of evidence include red $K'-L'$ compared to young free-floating objects \citep{K14}, the 1.28~\micron~Pa-$\beta$ emission line (\citealt{S07}; but also see \citealt{M07} and \citealt{L09}), and overluminosity in the $HST$ F606W flux \citep{MMB07}. The H$\alpha$ emission, along with optical continuum excess, were detected by \cite{Z14}. A relatively high accretion rate $\dot{M}\sim10^{-9.3}~M_\sun~\mathrm{yr}^{-1}$ was derived by modeling the continuum excess as a hot hydrogen slab \citep{Z14}. The dust mass in the disk was shown to be $<$0.04 $M_\earth$ \citep{M16}.

Recently, \cite{S16} measured the spin and the barycentric radial velocity (RV) of GQ Lup B using high-dispersion spectroscopy. They showed that compared to the giant planet $\beta$ Pic b, GQ Lup B spins slowly because it is still contracting and gaining angular momentum from its disk. Their new RV estimate, together with the astrometric monitoring in \cite{G14}, has provided constraints on its orbit. \cite{S16} also detected CO and $\mathrm{H}_2$O in GQ Lup B's atmosphere.

We list the properties of GQ Lup A and B in Table \ref{tab:prop_gqlup}.

\begin{deluxetable}{@{}lccl@{}}
\tablecaption{Properties of GQ Lup \label{tab:prop_gqlup}}
\tablehead{
\colhead{\hspace{-38pt}Parameter} &
\colhead{GQ Lup A} &
\colhead{GQ Lup B} &
\colhead{\hspace{-20pt}References} 
}
\startdata
Distance (pc)					& 	\hspace{-4.2pt}$\sim$150		& 	\hspace{-4.2pt}$\sim$150			& 	1, 2	\\
Separation (\arcsec)				& 	$\cdots$					& 	$0.721\pm0.003$				&	3	\\
PA (\degr)						& 	$\cdots$					& 	\hspace{-8.5pt}$277.6\pm 0.4$ 		&	3	\\
Age (Myr)						& 	2--5 						& 	2--5							&	4	\\
SpT							&	K7eV					&  	\hspace{2.8pt}L$1\pm1$			& 	5, 6	\\
$A_V$ (mag)					&	\hspace{3.8pt}$0.4\pm0.2$	&	$\cdots$ 						&	7	\\
log($L/L_\sun$) 				&	\hspace{3.8pt}$0.0\pm0.1$	& 	\hspace{1.4pt}$-2.47\pm0.28$	 	&	4, 6	\\
$T_\mathrm{eff}$ (K) 			&	\hspace{-4.7pt}$4300\pm50$	&	\hspace{3.8pt}$2400\pm100$		&	4, 6	\\
Radius 		  				&	$1.7\pm0.2~R_\sun$			&	$3.4\pm1.1~R_\mathrm{Jup}$		&	4, 8	\\
Mass						&	\hspace{3.8pt}$1.05\pm0.07~M_\sun$	& 	$\sim$10--40 $M_\mathrm{Jup}$	& 	4, 6, 9, 10, 11, 12	\\
log $\dot{M}$ ($M_\sun$~yr$^{-1}$)	&	$-$9 to $-$7				&	$-$12 to $-$9					&	3, 4, 13, 14, 15	\\
log $g$ 						&	$3.7\pm0.2$				&	$4.0\pm0.5$					&	4, 6	\\
Inclination	(\degr)				&	$27\pm5$					& 	$\cdots$						& 	16	\\
$v~\mathrm{sin}(i)$ (km $\mathrm{s}^{-1}$)		&	$5\pm1$		&	$5.3^{+0.9}_{-1.0}$	 			& 	4, 17	\\
Rotation Period	(d)				&	$8.45\pm0.20$				& 	$\cdots$						&	16
\enddata
\tablerefs{(1) \citealt{C00}. (2) \citealt{F02}. (3) This work. (4) \citealt{D12}. (5) \citealt{H77}. (6) \citealt{L09}. (7) \citealt{B01}. (8) GQ Lup B's radius is derived from the adopted $L$ and $T_\mathrm{eff}$. (9) \citealt{N08}. (10) \citealt{MMB07}. (11) \citealt{M07}. (12) \citealt{S07}. (13) \citealt{H09}. (14) \citealt{SD08}. (15) \citealt{Z14}. (16) \citealt{B07}.  (17) \citealt{S16}.}
\end{deluxetable}

\section{Methodology}
\subsection{ALMA 1.3 mm}
GQ Lup was observed with ALMA in Cycle 3 on UT 2015, November 1 with the Band 6 receiver and 41 12-m antennas reaching a maximum baseline of 14969.3 meters. Three of the 4 available basebands were configured for continuum observations to search for dust emission, each with 128 15.625 MHz channels for a total of 2 GHz continuum bandwidth, and centered at 233.0, 246.0, and 248.0 GHz. The final baseband was centered at 230.538 GHz with 3840 0.122 MHz (Hanning smoothed to a resolution of 0.244 MHz, or 0.32 km s$^{-1}$) channels to search for ${}^{12}$CO (2--1) emission from our targets. Scans on GQ Lup were interleaved with the phase calibrator QSO J1534-3526. The total on-source time was 11.19 minutes.

\begin{figure}
\centering
\includegraphics[angle=0,width=\columnwidth]{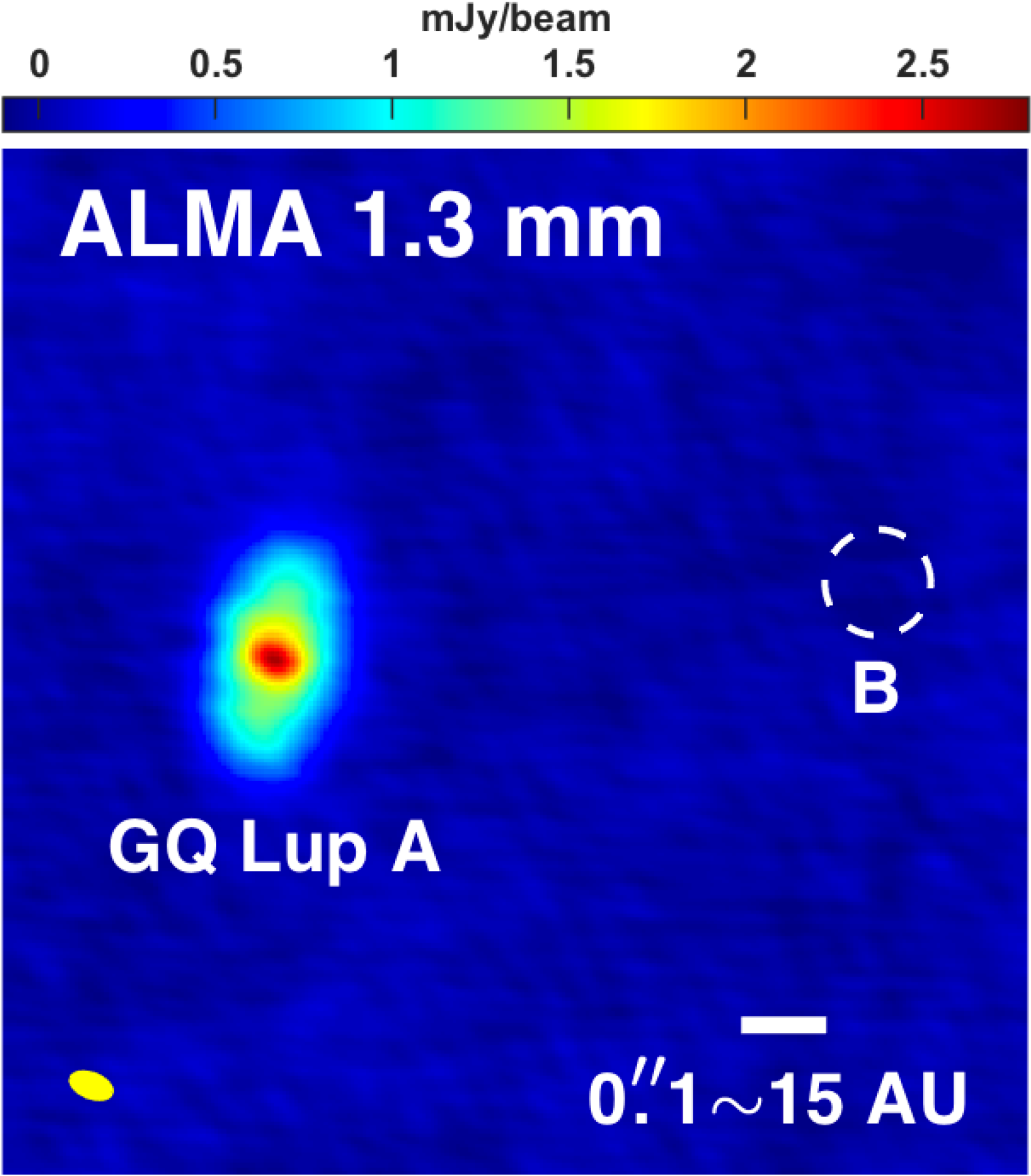}
\caption{ALMA 1.3 mm continuum map showing GQ Lup A's accretion disk. We did not detect B's disk (dashed circle). The $0\farcs054\times0\farcs031$ beam (8.1 AU $\times$ 4.7 AU at a distance of 150 pc) is shown as a yellow oval. North is up and east is left.}
\label{fig:alma_cont}
\end{figure}

The data were reduced in the standard way with the \texttt{CASA} software package, using the water vapor radiometry data and QSO J1534-3526 for gain calibration, QSO J1427-4206 for bandpass calibration, and QSO J1337-1257 for flux calibration. The calibrated data were Fourier inverted and deconvolved from the beam using the MSMFS-CLEAN algorithm \citep{RC11} with no frequency dependence, CLEAN components which are point sources and 1, 2, 4, and 8 times the size of the beam, and natural weighting for the best sensitivity. We produced a continuum map from all four basebands, excluding channels in the $-15$ to 15 km s$^{-1}$ range of the CO baseband to avoid contaminating our map with CO emission. The final 1.3 mm continuum map was produced after four iterations of phase-only self-calibration using a model produced from the CLEAN algorithm. We show the 1.3 mm continuum map in Figure \ref{fig:alma_cont}. The continuum map has an rms of 39 $\mu$Jy beam$^{-1}$, with a synthesized beam of $0\farcs054\times0\farcs031$ and position angle of $68\fdg7$.

\subsection{Sources in the ALMA Image}
\label{Section:alma_detections}
In the continuum map there is a clear detection of a source located at 15$^{\mathrm{h}}$49$^{\mathrm{m}}$12\fs09, $-$35\degr39\arcmin05\farcs43. Accounting for GQ Lup A's proper motion ($-15.1\pm2.8$ mas yr$^{-1}$, $-23.4\pm2.5$ mas yr$^{-1}$; \citealt{Z10}), this is coincident with its reported position at 15$^{\mathrm{h}}$49$^{\mathrm{m}}$12\fs10, $-$35\degr39\arcmin05\farcs12 (J2000). We measured a flux of $27.5\pm0.6$ mJy for this source.

As we know the position of GQ Lup B, we can search a smaller region of the image with a lower detection threshold for emission. Noise in our ALMA map is highly Gaussian, so we would expect 68\% of the peaks to be within 1$\sigma$ of zero, 95\% to be within 2$\sigma$, and so on. In a 0\farcs1 diameter region around the known position of GQ Lup B there are 4 beams, so we would expect $\sim$0.2 noise peaks above 2$\sigma$, but $\ll$1 noise peak above 3$\sigma$, so any peak above 3$\sigma$ is likely real. However, we did not detect any emission from GQ Lup B.

\subsection{GQ Lup B's Disk Mass}
\label{Section:diskmassupperlimit}
To place an upper limit on the disk mass of GQ Lup B, we inserted fake sources into our image in a 0\farcs1 diameter area around the known position of GQ Lup B, and used our source finding routine to search for them. We varied both the disk size and the flux of the input sources, and for each disk size/flux combination we calculated the percentage of the fake sources that were recovered in the map. For each disk size, we set the upper limit on disk flux to be the minimum flux for which 99.7\% ($3\sigma$) of the input fake sources were detected.

Flux upper limits can be converted into dust mass upper limits by assuming the dust is optically thin and using the standard prescription \citep{B90},
\begin{equation}
\label{eq:diskmass}
M_{\text{disk}} = \frac{D^2\, F_{\text{disk}}}{\kappa_{\nu}\, B_{\nu}(T)}.
\end{equation}
We used the standard assumption of a characteristic dust temperature of $T=20\,$K and a 1.3 mm dust opacity of $\kappa_{\nu} =$ 2.3 cm$^2$ g$^{-1}$. We used a distance of 150 pc to GQ Lup.

\begin{figure}
    \centering
    \includegraphics[width=\columnwidth]{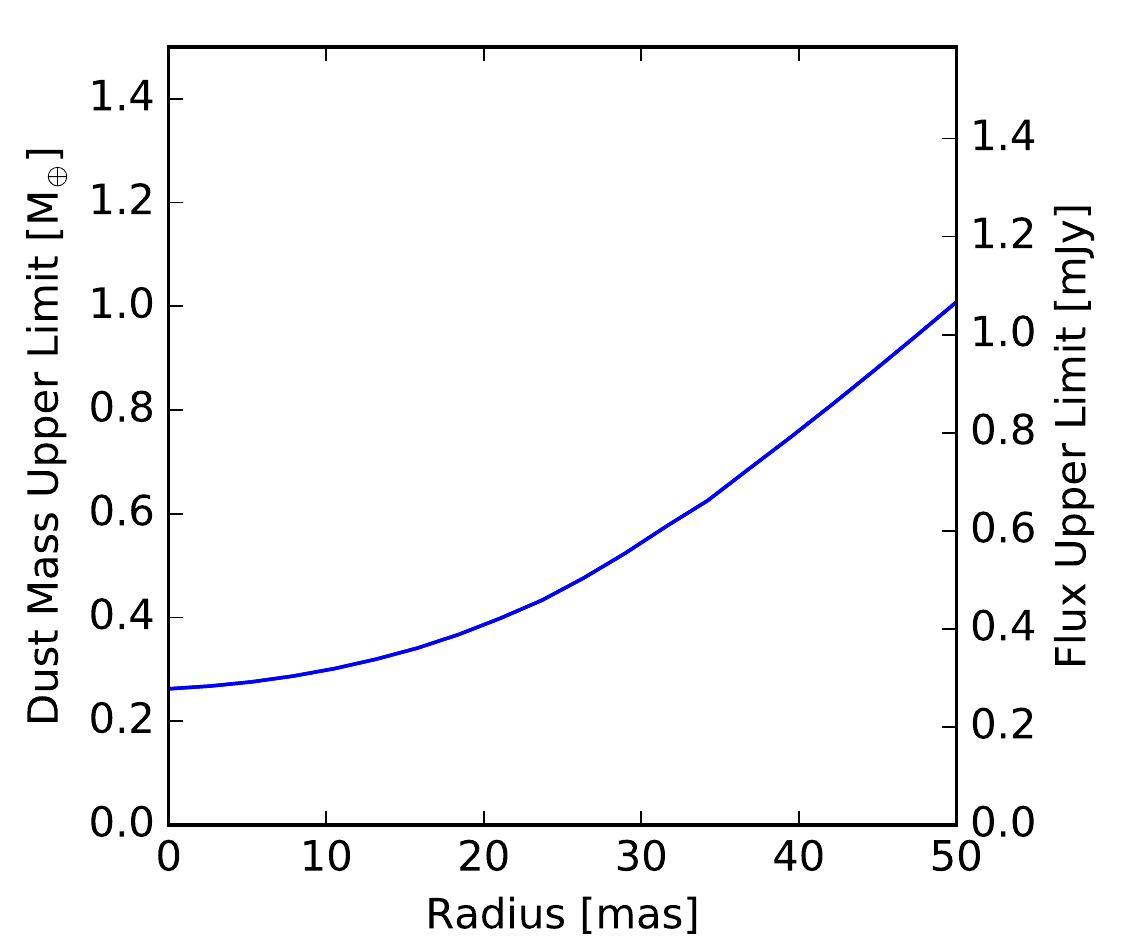}    
    \caption{The 3$\sigma$ upper limit on the dust mass of the GQ Lup B disk as a function of disk radius. The disk has a flux of $<$0.3 mJy corresponding to $<$0.25 $M_\earth$ of dust if it is a point source, or a flux of $<$1.1 mJy corresponding to $<$1 $M_\earth$ of dust if it reaches 1/3 Hill radius ($\sim$7.5 AU). We note that the recent 870 \micron~imaging by \cite{M16} placed a much stronger constraint on the dust mass, $<$0.04 $M_\earth$.}
    \label{fig:upper_limits}    
\end{figure}

In Figure \ref{fig:upper_limits} we show a plot of the upper limit on GQ Lup B's disk mass as a function of disk radius. The calculation ran from a point source disk to a disk with a radius of 50 mas. This radius corresponds to a third of the Hill radius for GQ Lup B, which is expected to be the upper limit on the size of the disk. This calculation also assumed that the projected separation of 0\farcs721 is equal to the semi-major axis of the companion's orbit. The Hill radius, and therefore the expected disk radius upper limit, would be larger if the system has a larger separation.

\subsection{GQ Lup A Disk Modeling}
As shown in Figure \ref{fig:alma_cont}, the disk around the primary star is strongly detected at 1.3 mm continuum. We fit the visibility data for GQ Lup A with a series of two disk models to constrain the parameters of the disk.

We modeled the disk with a detailed radiative transfer modeling scheme, using the RADMC-3D code \citep{Dullemond12} to produce disk models. Our model included a central protostar with a temperature of 4300 K and luminosity of 1 $L_{\odot}$, consistent with the measured values (see Table \ref{tab:prop_gqlup}). It also included a dusty protoplanetary disk, for which we used two different density prescriptions. Model A used the density profile of a flared power-law disk,
\begin{equation}
\Sigma = \Sigma_0 \left(\frac{R}{1 \text{ AU}}\right)^{-\gamma}, \text{ and}
\end{equation}
\begin{equation}
\rho = \frac{\Sigma(R)}{\sqrt{2\pi} \, h(R)} \, \exp\left(-\frac{1}{2}\left[\frac{z}{h(R)}\right]^2\right), \text{  with}
\end{equation}
\begin{equation}
h(R) = h_0 \left(\frac{R}{1 \text{ AU}}\right)^{\beta}.
\end{equation}
We allowed the disk mass ($M_{\text{disk}}$), inner and outer disk radii ($R_{\text{in}}$ and $R_{\text{disk}}$), the surface density index ($\gamma$), scale height index ($\beta$), and scale height at 1 AU ($h_0$), to vary as free parameters. Model B used the surface density profile of a flared accretion disk with a radial power-law distribution of viscosity \citep[e.g.,][]{LBP74},
\begin{equation}
\Sigma = \Sigma_0 \left(\frac{R}{R_c}\right)^{-\gamma} \exp\left[-\left(\frac{R}{R_c}\right)^{2-\gamma}\right].
\end{equation}
The parameters for Model B were the same as those of Model A, with the exception of the critical radius $R_{c}$, beyond which the surface density drops exponentially. This replaced the disk radius, beyond which the density drops to zero in Model A. We also supplied dust opacities to the models. We assumed that dust grains are 70\% astronomical silicate and 30\% graphite \citep{WD01} following a power-law distribution of dust grain sizes, with a minimum size of 5 nm, a maximum size of 3 mm, and a power law exponent of $-$3.5. This produced a 1.3 mm opacity of 2.25 cm$^{2}$ g$^{-1}$, in good agreement with the typical value assumed for disk mass calculations.

\begin{figure*}[t]
    \centering
    \includegraphics[width=\linewidth]{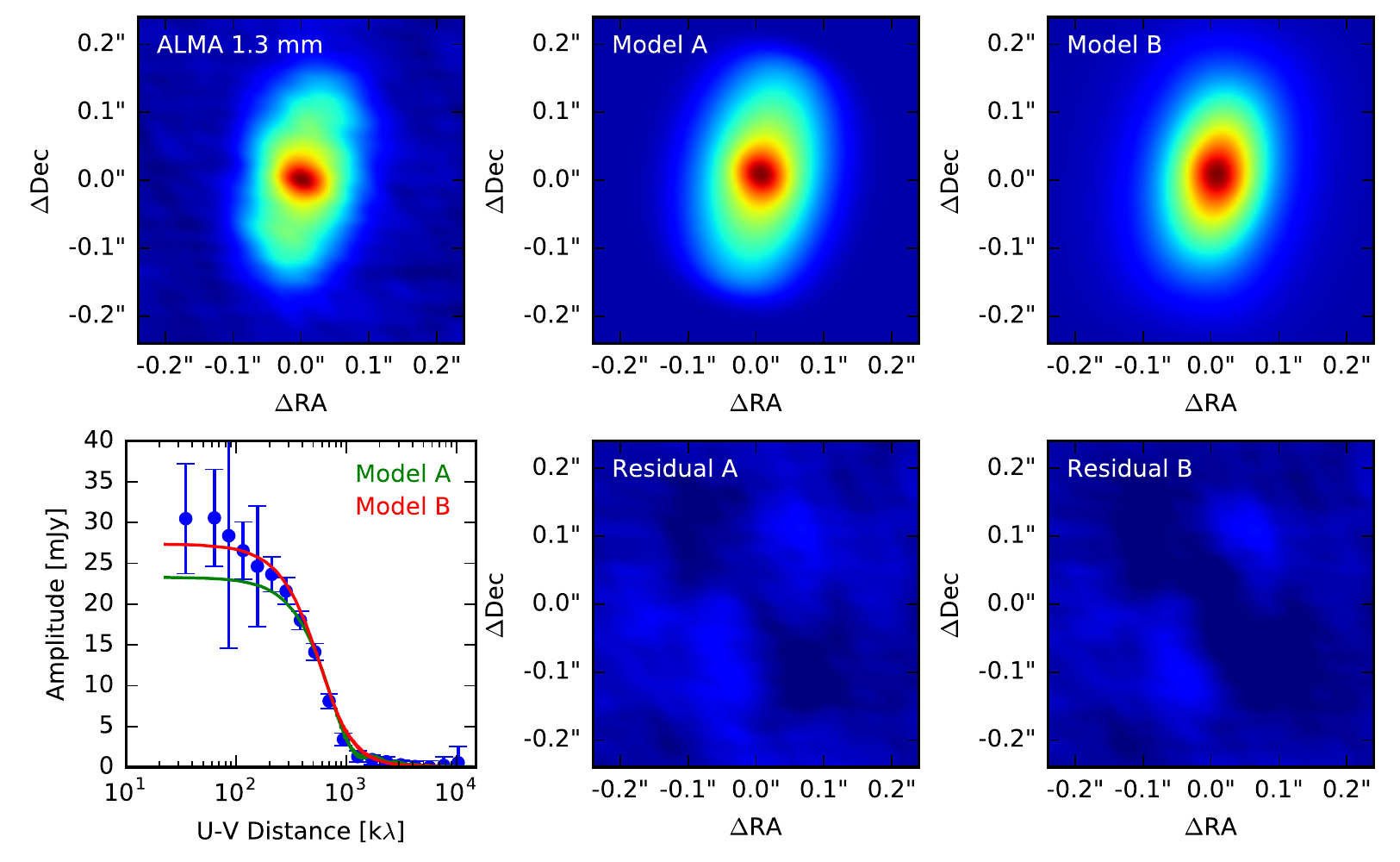}
    \caption{The best-fit models to the GQ Lup A disk. In the first column we show the ALMA 1.3 mm dust continuum image and averaged 1D visibilities. The colored lines show the best fits for Models A and B. In subsequent columns we show the best-fit image and residual for each model.}
    \label{fig:models}
\end{figure*}

We used RADMC-3D to calculate the temperature throughout the density distribution. Following this we produced synthetic images with raytracing of the protostar model, and Fourier transformed the images to produce model 1.3 mm visibility profiles. We fit these models directly to the visibility data using the Markov Chain Monte Carlo code \texttt{emcee} \citep{FM13}. \texttt{emcee} uses an affine-invariant MCMC ensemble sampler, which employs a series of walkers that step through parameter space and converge on the best fit. We positioned 200 walkers throughout a large region of parameter space and allowed them to move towards regions of lower $\chi^2$. We show the best fit parameters for each model in Table \ref{tab:disk_parameters}, and images, residuals, and visibilities for the best-fit models in Figure \ref{fig:models}.

\subsection{MagAO H$\alpha$ Photometry}
GQ Lup was observed on UT 2015, April 16 in the simultaneous differential imaging (SDI) mode \citep{C14} of the VisAO camera ($8\arcsec\times8\arcsec$ field of view, 79 mas plate scale; \citealt{C13}). The rotator was off to facilitate angular differential imaging (ADI) \citep{M06}. Weather was photometric with low ground level winds. Seeing varied from 0\farcs7 to 0\farcs8. AO parameters and exposure time are listed in Table \ref{tab:MagAO_obs}. Images in the H$\alpha$ (656 nm; $\Delta\lambda$ = 6.3 nm) and continuum (643 nm; $\Delta\lambda$ = 6.1 nm) channels were separated, dark subtracted, registered and centered, and then the median radial profile of each image was subtracted from itself. 

After these basic reduction steps, we then proceeded to employ the KLIP algorithm \citep{S12} for PSF subtraction with ADI. Our implementation of the ADI+KLIP algorithm allows for the selection of many parameters, including the size of the search region, a minimum rotation requirement, and the number of modes \citep{Males14}. The following sequence of steps was taken to find a set of signal-to-noise (S/N) optimizing reduction parameters in an unbiased way.

\begin{figure*}[t]
\centering
\includegraphics[angle=0,width=0.33\linewidth]{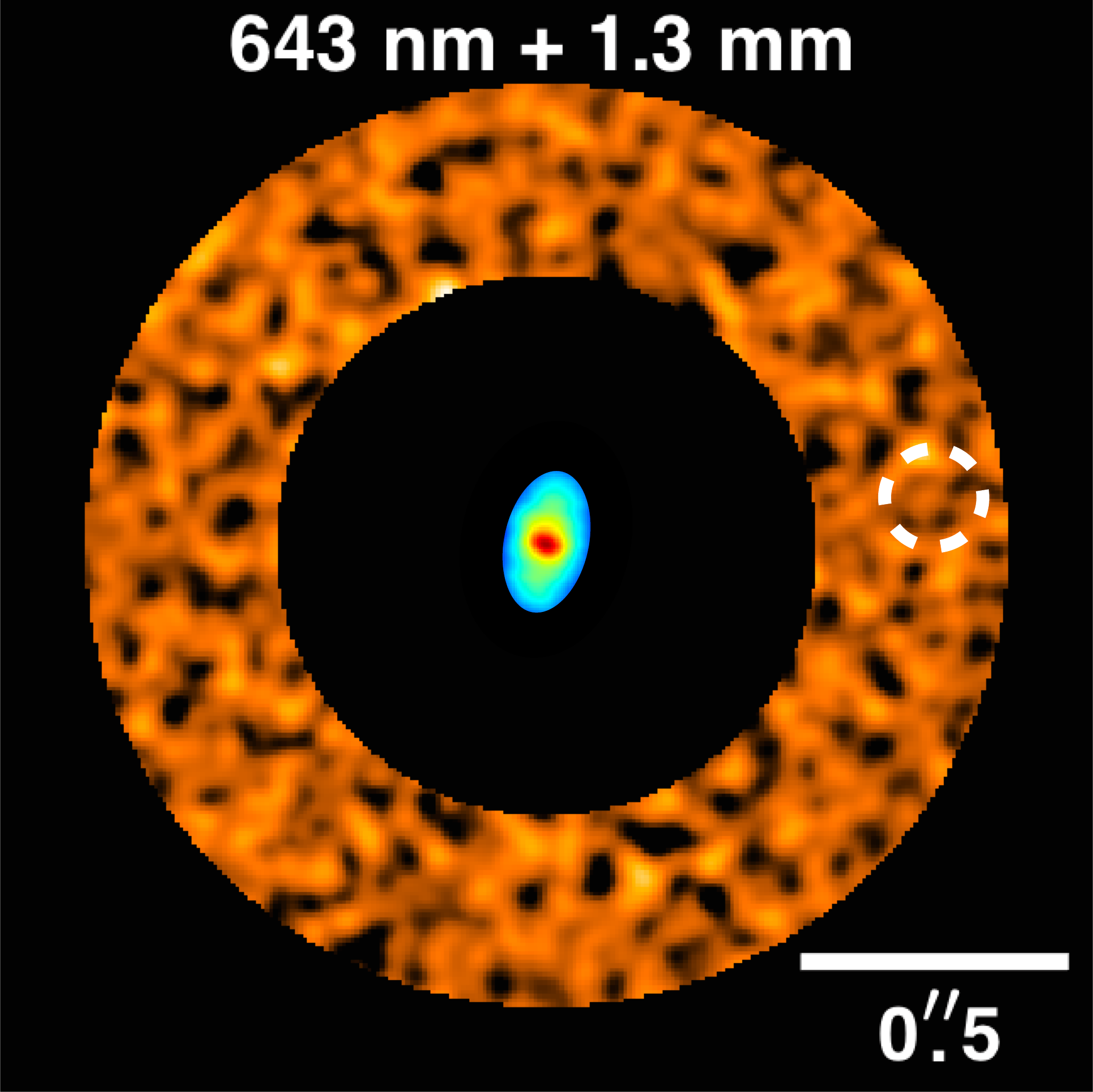}
\includegraphics[angle=0,width=0.33\linewidth]{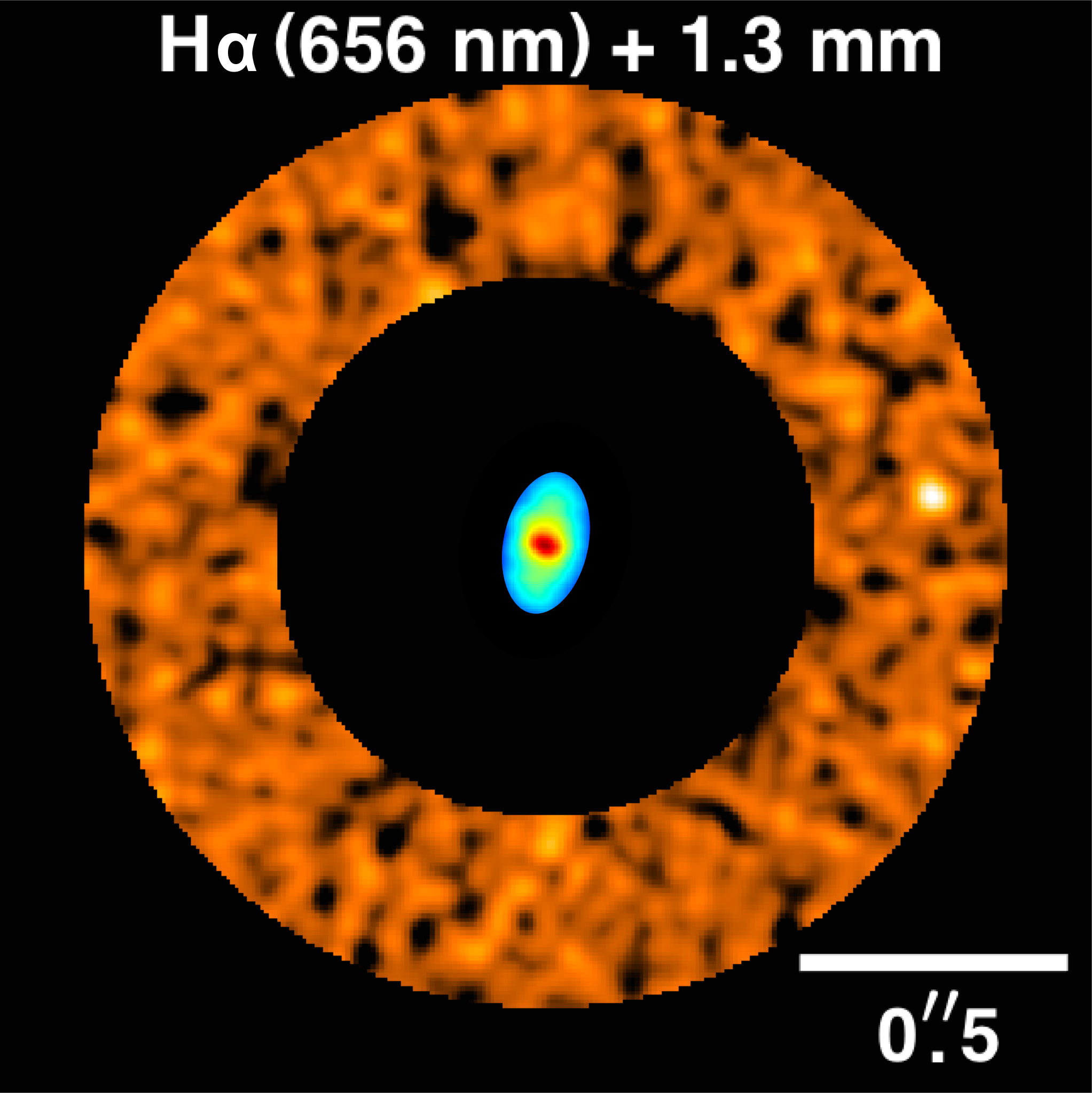}\\
\includegraphics[angle=0,width=0.33\linewidth]{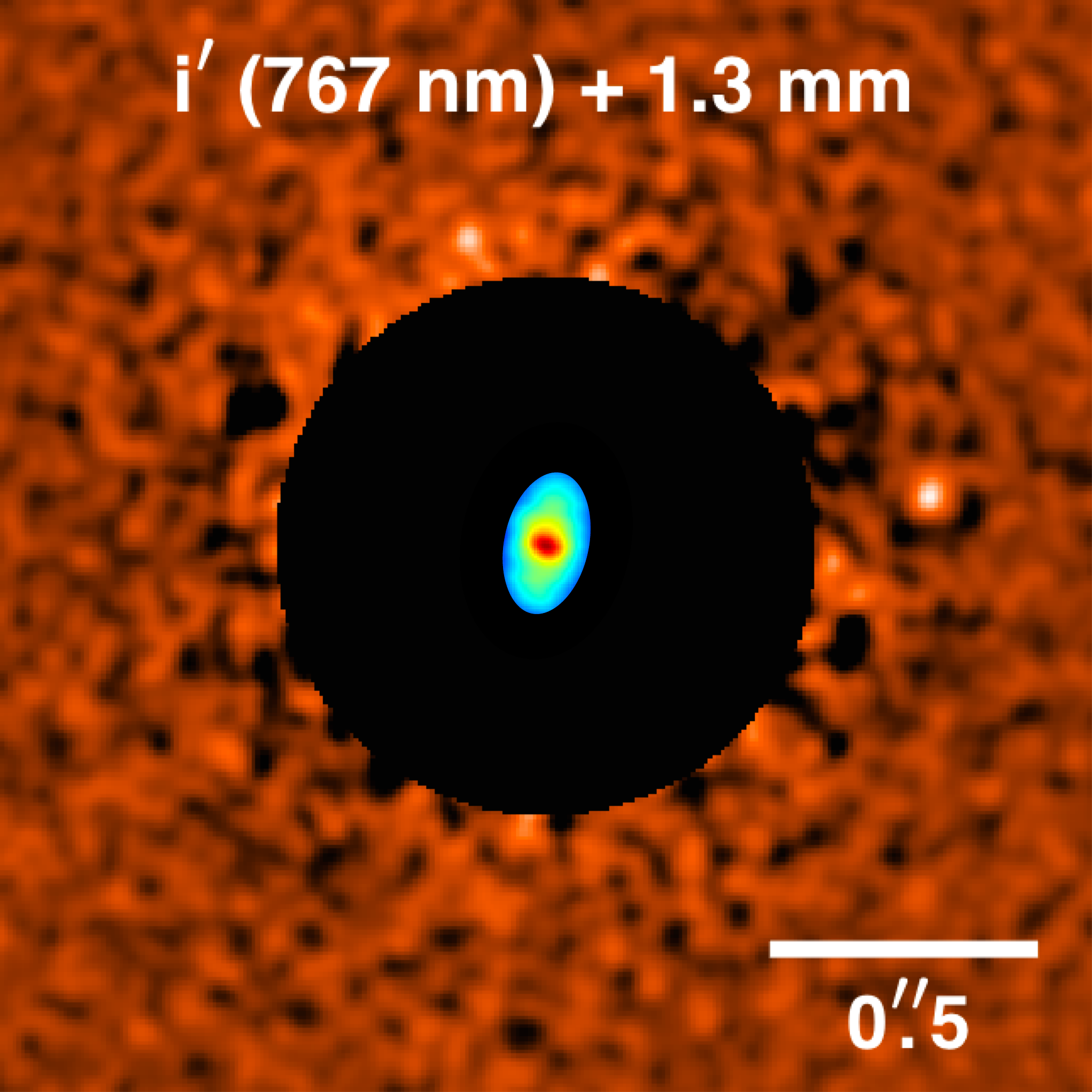}
\includegraphics[angle=0,width=0.33\linewidth]{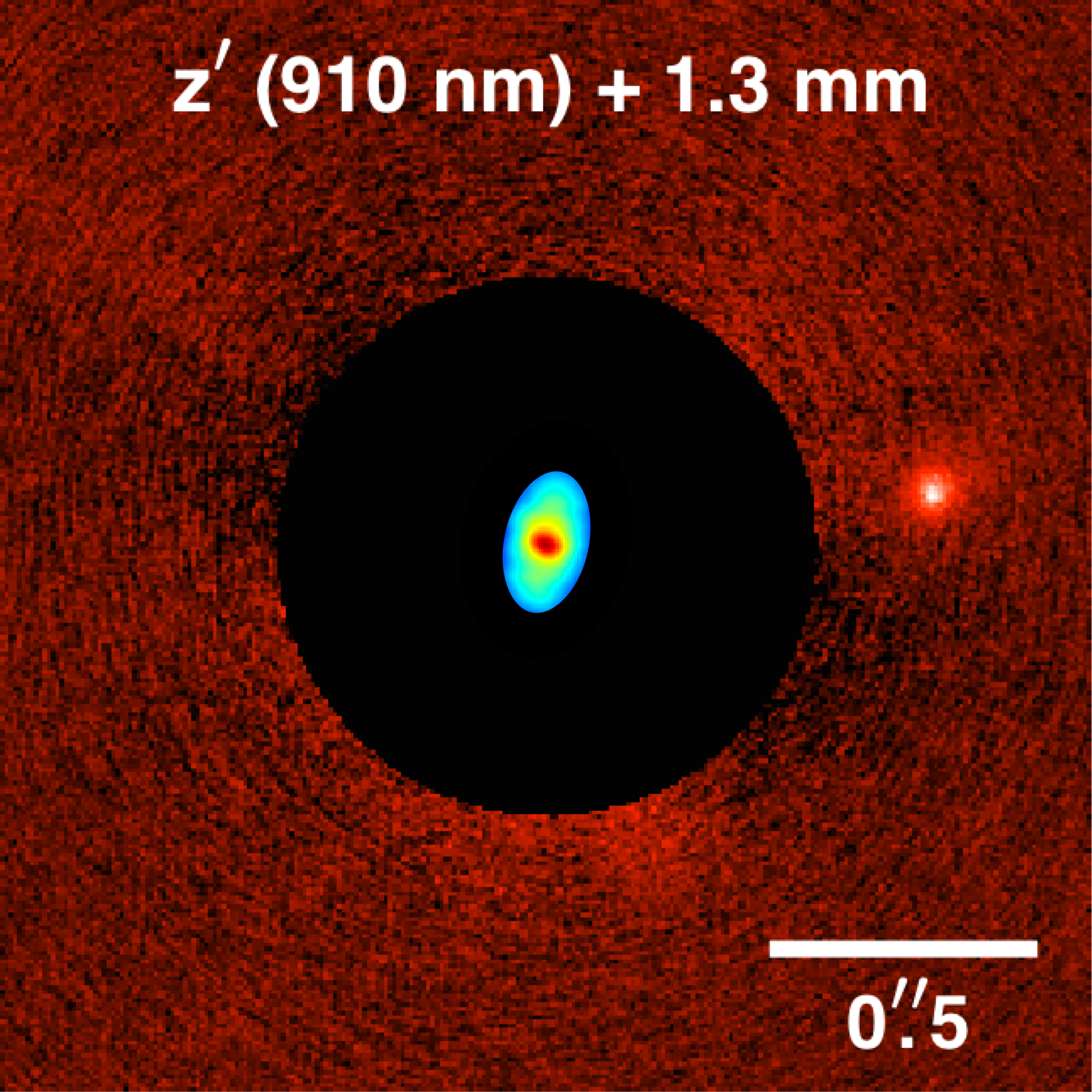}
\includegraphics[angle=0,width=0.33\linewidth]{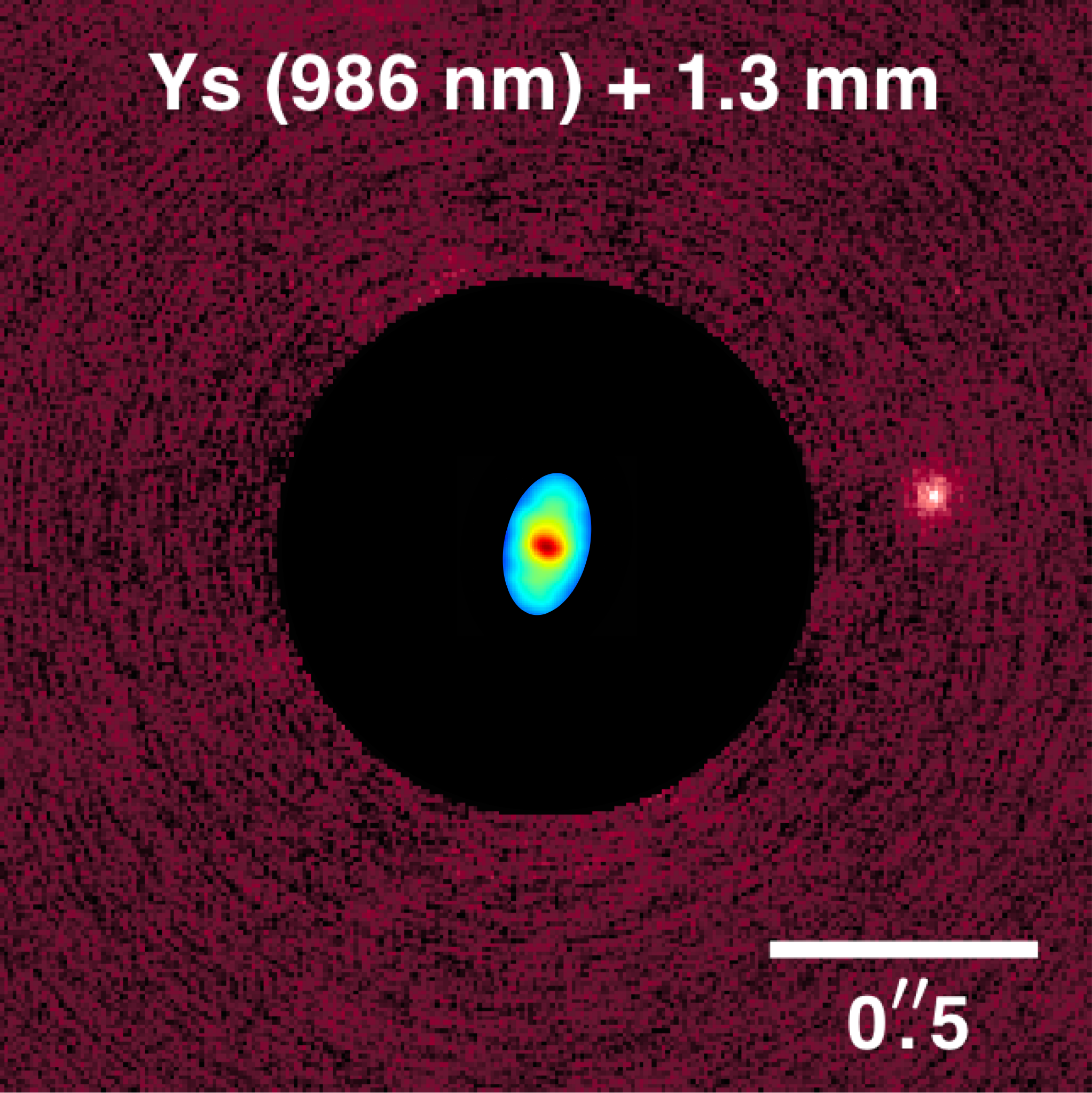}
\caption{GQ Lup B in MagAO filters. We mask out the central 0\farcs5 and overlay the ALMA 1.3 mm disk image. GQ Lup B is glowing at H$\alpha$ but not the 643 nm continuum, indicating active accretion. North is up and east is left.}
\label{fig:MagAO_ALMA}
\end{figure*}

First, an initial reduction was carried out to locate the companion. This employed a search annulus from 50 to 150 pixels (0\farcs4 to 1\farcs2), minimum of 2.5 pixels of rotation at the inner edge between the image being reduced and basis images, and modes ranging from 5 to 30. After PSF subtraction, images were de-rotated and the final image was formed as the 5$\sigma$-clipped mean. The final image was then unsharp-masked with a 20-pixel Gaussian kernel, and then smoothed with a 5-pixel Gaussian. Unsharp masking acts a high-pass filter, removing the stellar halo and the residual low-spatial frequency noise remaining after PSF subtraction. Given the low quality correction which yielded an FWHM much larger than $\lambda/D$, the PSF was oversampled by the diffraction limited plate scale of the VisAO camera. Gaussian smoothing (low-pass filter) hence smooths pixel-to-pixel noise. The companion was readily seen in the H$\alpha$ channel with S/N $\sim$ 5 using 8 modes, but no detection was evident in the continuum channel (upper panels of Figure \ref{fig:MagAO_ALMA}).

Next, we injected negative planets \citep{Bonnefoy11} with the same search region, using 8 modes to form an initial estimate of the planet flux, finding $\Delta$H$\alpha$ $\sim$ 8.7 mag as the planet brightness which minimized the standard deviation in an aperture with radius of 1 FWHM at the location of the companion.  

It is difficult to estimate uncertainties using the negative planet technique (e.g., \citealt{M15}), and optimizing the reduction parameters on the companion itself risks biasing results due to speckles. To address these issues, we conducted a grid search over the parameters, testing on an ensemble of positive fake planets injected at the same separation but at a range of position angles. To conduct this search in a reasonable amount of time we employed the ``Findr'' distributed computing (cloud-based) data reduction system \citep{HB16}. In all trials, a negative planet with the above estimated brightness was injected at the location of the detection to avoid cross-talk. Planet injection was performed on the dark-subtracted/registered/centered images, and then the radial profile was subtracted. The parameters tested are given in Table \ref{tab:Halpha_redux}, and the KLIP algorithm was applied for each possible combination. The final combined image at each combination was filtered as above. The optimum parameters were determined as those which maximized the mean S/N on the ensemble of positive fake planets injected at 8.7 mag brightness. The flux of the companion was then determined by comparing its photometry when reduced with those optimum parameters to the positive fake planet results. Uncertainties were estimated from the standard deviation of the results from the fake planets.

At the optimum parameters, GQ Lup B was detected in H$\alpha$ at S/N $\sim$ 6.3 with a contrast of 8.60 $\pm$ 0.16 mag. For comparison, the H$\alpha$ contrast in \cite{Z14} is $\sim$7.1 mag (Zhou, Y. 2016, private communication). The contrast at 643 nm is $>$8.81 mag (3$\sigma$ upper limit). We also found that GQ Lup A's H$\alpha$ flux is $\sim$0.60 mag brighter than its 643 nm continuum due to active accretion.

\subsection{Derivation of Accretion Rates from H$\alpha$ Fluxes}
We derived the mass accretion rates for GQ Lup A and B following \cite{R12} and \cite{C14}. In brief, we computed the H$\alpha$ line luminosity $L_{\mathrm{H}\alpha}$, used it to derive the accretion luminosity $L_\mathrm{acc}$, and applied the energy equation $L_\mathrm{acc} \propto GM_\star \dot{M}/R_\star$ to determine $\dot{M}$.

Since we did not observe a standard star at $R$ band, we estimated GQ Lup A's average $R$ brightness to be 11.0 mag by averaging literature fluxes in \cite{SN78}, \cite{B82}, \cite{KF85}, and \cite{C92}, with no attempt to homogenize photometric systems. The uncertainty was taken to be 0.7 mag as \cite{B07} found that GQ Lup A's $R$ flux can change by 1.4 mag over its 8.45-day rotation period. To recover the true $L_{\mathrm{H}\alpha}$, we also have to correct for dust extinction, which is $A_V\sim0.4$ mag to the star \citep{B01} but unknown to the companion. As a result, our $\dot{M}$ estimate for GQ Lup B should be considered a lower limit. 

We thus estimated $L_{\mathrm{H}\alpha} \sim 10^{-2.3}$ to $10^{-1.8}~L_\sun$ for A, and $\sim10^{-5.9}$ to $10^{-5.4}~L_\sun$ for B. Substituting $L_{\mathrm{H}\alpha}$ into the empirical relation, log$(L_\mathrm{acc}) =  2.99 + 1.49~\times$ log$(L_{\mathrm{H}\alpha})$, in \cite{R12}, we found $L_\mathrm{acc} \sim 0.3$ to 2.3 $L_\sun$ for A, and $\sim 1.4\times10^{-6}$ to $9.7\times10^{-6}~L_\sun$ for B. The resulting accretion rates for GQ Lup A and B are $\dot{M}\sim10^{-8}$ to $10^{-7}~M_\sun~\mathrm{yr}^{-1}$, and $\sim10^{-12}$ to $10^{-11}~M_\sun~\mathrm{yr}^{-1}$, respectively. Our measurement for A is similar to previous results, $\dot{M}\sim10^{-9}$ to $10^{-7}$ $M_\sun~\mathrm{yr}^{-1}$ \citep{SD08,H09,D12}. On the other hand, for the companion we obtained a lower value compared to $10^{-9.3}~M_\sun~\mathrm{yr}^{-1}$ in \cite{Z14}. Possible causes include the unknown dust extinction or an inactive period of accretion during the time of our observations. Our $\dot{M}$ estimates and literature values are also shown in Table \ref{tab:prop_gqlup}.

\begin{figure*}[t]
\includegraphics[angle=0,width=0.494\linewidth]{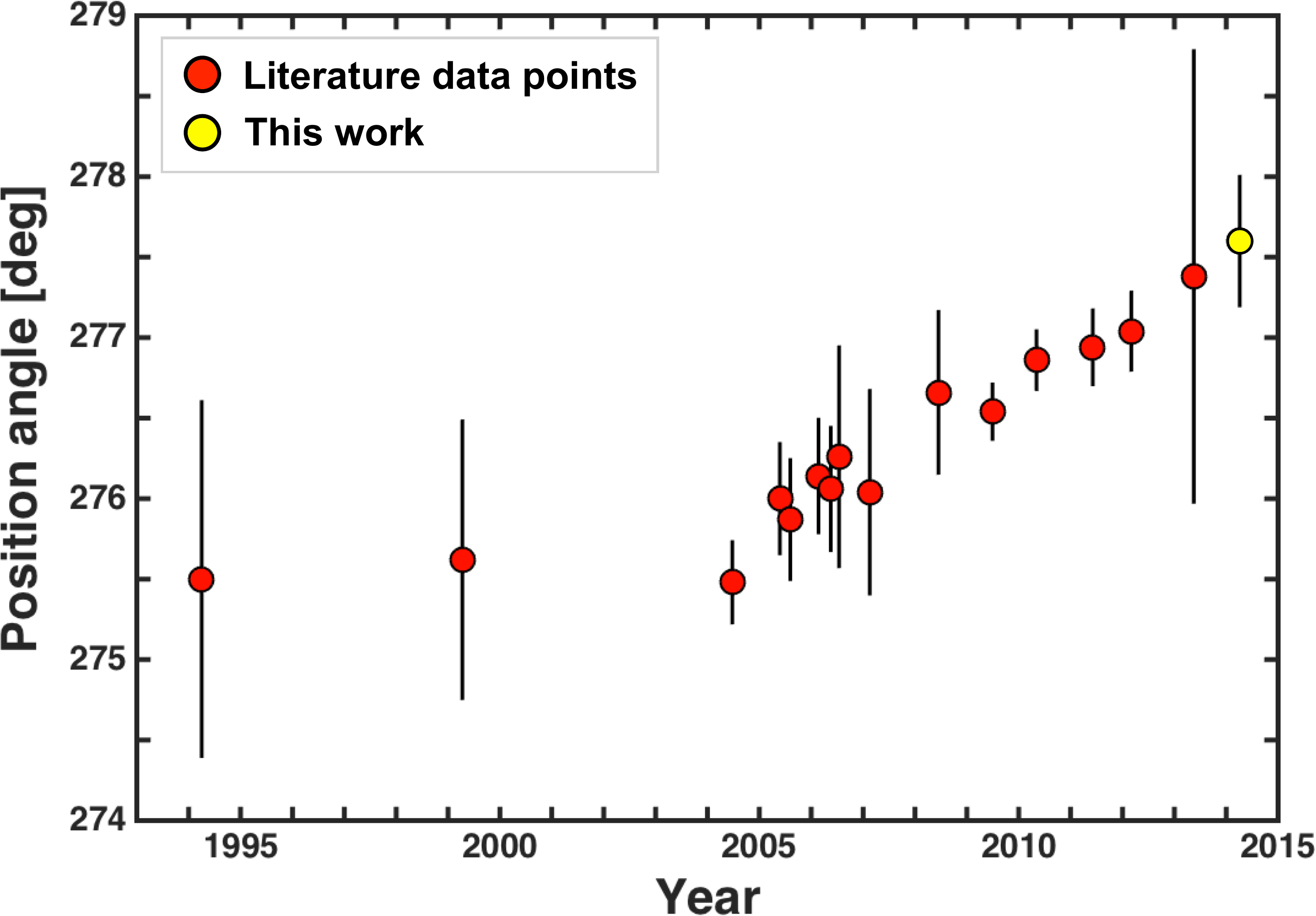}
\includegraphics[angle=0,width=0.495\linewidth]{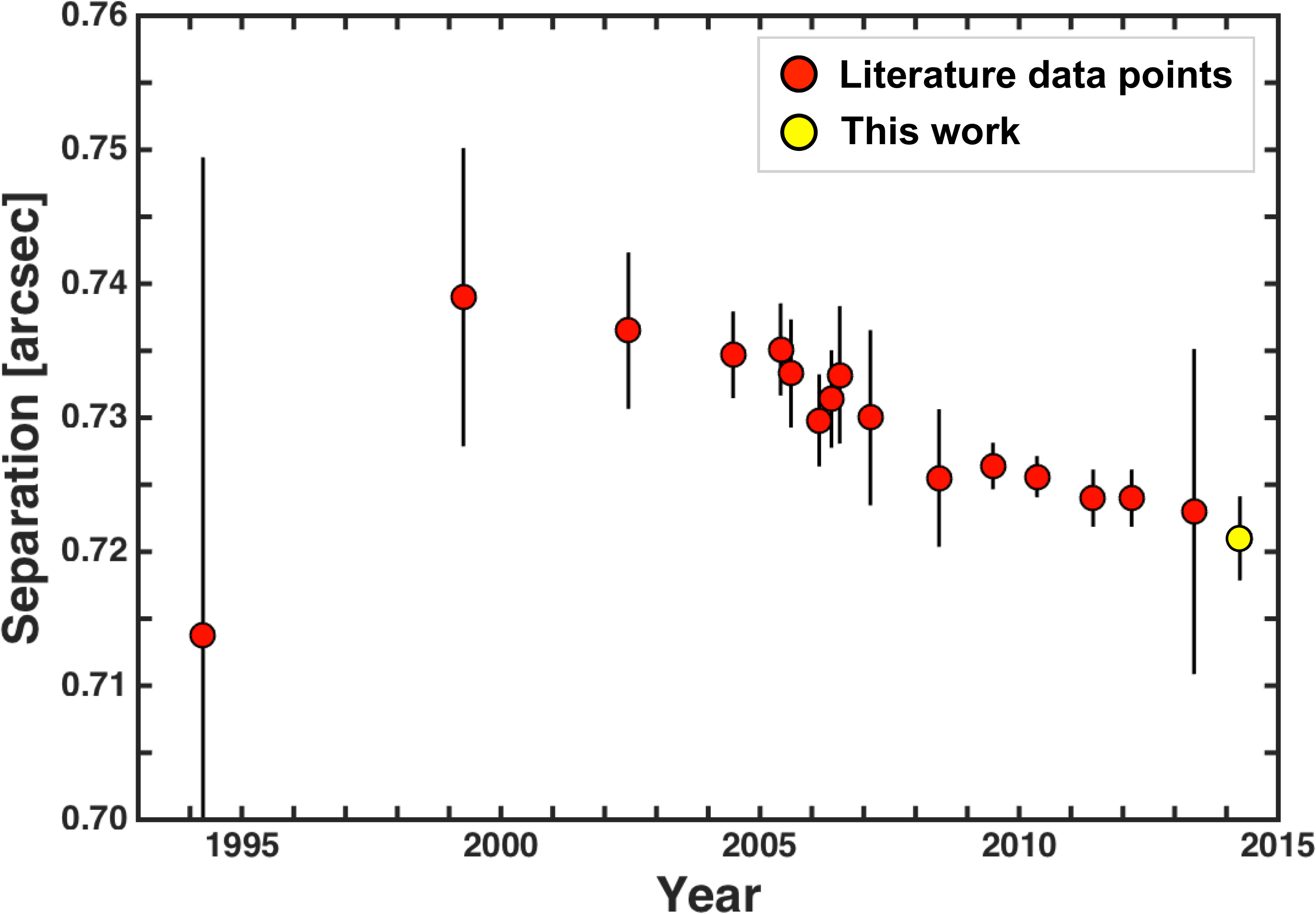}
\caption{Orbital motion of GQ Lup B over the last two decades. Existing data points are compiled from \cite{J06}, \cite{N05,N08}, \cite{G14}, and \cite{U17}. Figure adapted from \cite{G14}.}
\label{fig:astrometry}
\end{figure*}

\subsection{MagAO $i'$, $z'$, $Y_S$ Photometry}
We observed GQ Lup at broad-band filters $i'$ (0.77\micron, $\Delta\lambda$ = 0.13~\micron), $z'$~(0.91~\micron; $\Delta\lambda$ = 0.12~\micron), and $Y_S$ (0.98~\micron; $\Delta\lambda$ = 0.09~\micron) on UT 2014, April 5. Weather was partially cloudy with $\sim$1\farcs3 seeing. Data reduction and photometry were carried out with IRAF\footnotemark[1]\footnotetext[1]{IRAF is distributed by the National Optical Astronomy Observatories, which are operated by the Association of Universities for Research in Astronomy, Inc., under cooperative agreement with the National Science Foundation.}(\citealt{T86}, \citeyear{T93}) and MATLAB. Raw data were dark-subtracted, registered, de-rotated, and median-combined. We then subtracted the median radial profile of the combined image from itself. For $i'$, we further filtered out the residual using a 15-pixel Gaussian kernel, and smoothed the resulting high-pass filtered image with a 6-pixel Gaussian to better bring out the companion. Our $i'$, $z'$, and $Y_S$ detections of GQ Lup B are also shown in Figure \ref{fig:MagAO_ALMA}.

We estimated GQ Lup B's fluxes by injecting fake planets. For $z'$ and $Y_S$, the primary star in the unsaturated data was used to create fake planets. For $i'$, since we did not acquire any unsaturated data, we estimated the peak height of the primary star in the saturated data from its beam splitter optical ghost \citep{Males16}. Then, we flagged the saturated core, assigned the peak value to the center, and fit a two-component Gaussian to create a PSF template. To compensate for flux loss from data reduction, including filtering, we injected fake planets at the same separation from the star but with different position angles, and repeated the data reduction procedures. We found that throughputs were $>$98\% for $z'$ and $Y_S$, but only $\sim$32\% for $i'$ as it involved more aggressive spatial filtering and so higher losses of low spatial frequency flux. Compensating for these flux losses, we derived contrasts of $8.13\pm0.23$ mag, $6.63\pm0.05$ mag, and $6.45\pm0.05$ mag for $i'$, $z'$, and $Y_S$, respectively.

To perform absolute photometry, we compared GQ Lup A to the standard star GJ 440. We used an 80-pixel aperture to include most of the flux, and adopted an 8\% photometric uncertainty recommended for absolute photometry with the VisAO camera by \cite{Males16}. Finally, we obtained $i'=10.76\pm0.08$ mag, $z'=9.77\pm0.08$ mag, and $Y_S=9.43\pm0.08$ mag for GQ Lup A, and $i'=18.89\pm0.24$ mag, $z'=16.40\pm0.10$ mag, and $Y_S=15.88\pm0.10$ mag for GQ Lup B. For comparison, our $z'$ flux is similar to the F850LP flux of 16.2 mag in \cite{Z14}, but our $i'$ flux is $\sim$1 mag fainter than their F775W flux of 17.8 mag. It is possible that GQ Lup B was in a quiescent accretion state during our observations. Table \ref{tab:MagAO_photometry} summarizes our photometric measurements. 

\subsection{MagAO Astrometry of GQ Lup B}
Following the calibrations in \cite{C13} and \cite{Males14}, in our MagAO data we found that GQ Lup B is 0\farcs721 $\pm$ 0\farcs003 from its host star, with a position angle of 277\fdg6 $\pm$ 0\fdg4. Figure \ref{fig:astrometry} shows the astrometric monitoring in the last $\sim$20 years, and GQ Lup B's orbital motion is  evident. Our results are consistent with the trends derived by \cite{G14}, $\Delta\rho\sim-1.4$ mas yr$^{-1}$ and $\Delta$PA $\sim+0\fdg16$ yr$^{-1}$. Table \ref{tab:prop_gqlup} also lists our astrometric measurements.

\begin{deluxetable}{@{}lcc@{}}
\tablecaption{GQ Lup A Disk Properties\label{tab:disk_parameters}}
\tablehead{
\colhead{Parameter} &
\colhead{Model A} &
\colhead{Model B} 
}
\startdata
Dust Mass ($M_\earth$) & $5.9~\pm~1.0$ & $5.5~\pm~0.8$ \\
Total Mass\tablenotemark{\dag} ($M_\earth$) & $77.2~\pm~8.4$ & $76.8~\pm~8.3$ \\
Inner Radius (AU) & $1.5~\pm~0.8$ & $1.7~\pm~1.1$ \\
Radius (AU)	& $23.8~\pm~1.6$ & $19.5~\pm~1.4$ \\
$h_{0}$ & $0.084~\pm~0.065$ & $0.075~\pm~0.039$ \\
$\gamma$ & $0.10~\pm~0.22$ & $-0.21~\pm~0.20$ \\
$\beta$ & $1.26~\pm~0.19$ & $1.45~\pm~0.25$ \\
Inclination	(\degr) & $56.2~\pm~4.8$ & $55.3~\pm~6.0$ \\
PA (\degr) & $348.8~\pm~4.8$ & $348.6~\pm~5.0$ 
\enddata
\tablenotetext{$\dagger$}{Total mass is calculated by adding our dust mass to the gas mass of 71.3 $\pm$ 8.3 $M_\earth$ measured by \cite{M16}.}
\label{tab:disk_parameters}
\end{deluxetable}

\begin{deluxetable}{@{}lcccc@{}}
\tablecaption{MagAO Observations \label{tab:MagAO_obs}}
\tablehead{
\colhead{Filter} &
\colhead{AO speed} &
\colhead{AO modes} &
\colhead{$t_{\rm sat}$} &
\colhead{$t_{\rm unsat}$} 
}
\startdata
643 nm		&	300 Hz	&	120	&	$\cdots$			&	25~s $\times$ 143	\\
656 nm (H$\alpha$) 	& 	300 Hz 	&	120	&	$\cdots$			& 	25~s $\times$ 143	\\
$i'$			& 	625 Hz 	&	120	&	10~s $\times$ 20	& 	$\cdots$ 			\\
$z'$			& 	625 Hz 	&	120	&	5~s $\times$ 13	& 	0.283~s $\times$ 54	\\
$Y_S$		& 	625 Hz	&	120	&	$\cdots$			&	15~s $\times$ 14
\enddata
\end{deluxetable}

\begin{deluxetable*}{@{}lccc@{}}
\tablecaption{Parameters of H$\alpha$ KLIP Reduction\label{tab:Halpha_redux}}
\tablehead{
\colhead{Parameter} &
\colhead{Grid} &
\colhead{Optimum} &
\colhead{Notes} 
}
\startdata
Minimum radius of region (pixel)	&   40--80, steps of 10           	 	&   50		& \\
Maximum radius of region (pixel) 	&   110--140, steps of 10          	 	&   110		& \\
Minimum rotation (pixel)         		&   0.0, 0.25, 0.5, 1.0, 2.0       	 	&   1.0		& \\
Number of modes                   		&   2--20, steps of 2              	 	&   6			& \\
Fake Planet PA                    		&   7.15--357.15, steps of 10       	&   $\cdots$	& 33 total, skipped $\pm$10\degr~from planet \\
Fake Planet Contrast             		&   1.65, 3.30, $4.95 \times 10^{-4}$ 	&   $\cdots$ 	& $\pm$50\% from negative planet result
\enddata
\end{deluxetable*}

\begin{deluxetable}{@{}lccc@{}}
\tablecaption{MagAO Photometry \label{tab:MagAO_photometry}}
\tablehead{
\colhead{Contrast/Filter} &
\colhead{GQ Lup A} &
\colhead{	} &
\colhead{GQ Lup B}
}
\startdata
$\Delta$643 nm\tablenotemark{\dag}		&					&	 	$>8.81$				&	\\
$\Delta$H$\alpha$					&					&	 	$8.60\pm0.16$			&	\\
$i'$						&	\hspace{-4.2pt}$10.76\pm0.08$	&	&	$18.89\pm0.24$	\\
$z'$						&	$9.77\pm0.08$		&	&	$16.40\pm0.10$	\\
$Y_S$					&	$9.43\pm0.08$		&	&	$15.88\pm0.10$ 
\enddata
\tablenotetext{$\dagger$}{3$\sigma$ upper limit.}
\end{deluxetable}

\section{Results}
\subsection{GQ Lup A's Disk}
We list the best-fit parameters for our modeling of the GQ Lup A disk in Table \ref{tab:disk_parameters}, and show our best-fit models in Figure \ref{fig:models}. Both models match the data well, and produce similar best-fit parameter values. We find that GQ Lup A's disk has a radius of $\sim$22 AU, an inclination angle of $\sim$56\degr, and a position angle of $\sim$349\degr. The mass in dust in the disk is $\sim$6 $M_\earth$, lower than $\sim$9.5 $M_\earth$ found by \cite{D10} and $\sim$15 $M_\earth$ found by \cite{M16}. This difference in dust mass likely comes from the adopted temperature profiles in the disk. In this study, we use radiative transfer to calculate the local temperature. Alternatively, if we assume a constant temperature of 20 K throughout the entire disk and use Equation \ref{eq:diskmass} to calculate the dust mass from the measured flux of 27.5 mJy, we obtain a higher value of $\sim$18 $M_\earth$.

We find that the disk size we measure ($R\sim22$ AU) is smaller than the size measured by \citet{M16} ($R\sim30$ AU from 870 \micron~continuum, and $R\sim46.5$ AU from CO (3--2) emission). This may be because dust grain growth is expected to occur preferentially in the inner disk, where densities are higher, and radial drift will tend to concentrate large particles at smaller radii. Our 1.3 mm map is more sensitive to large dust grains than the 870 \micron~map in \citet{M16}, so we may expect to measure a smaller radius at longer wavelengths.

Our map does not show any structures in the GQ Lup A disk such as holes or gaps, which can be the signposts of additional companions. The best-fit disk models also have very small inner disk radii, of $\sim$1.6 AU. This is consistent with no inner clearing, because the resolution of our observations does not allow us to well constrain the inner radius below $\sim$4.5 AU. Since there is unlikely any gaps or companions hidden within the disk, GQ Lup B was probably not scattered to its current orbit, but instead formed in-situ like binary stars, as we discuss more in Section \ref{discussion}.

Our models show that the surface density profile of the disk can be very flat. The flat profile is similar to some brown dwarf disks in $\rho$ Ophiuchus \citep{T16}, but in contrast to brown dwarf disks in Taurus \citep{R14}, which have rather steep profiles and smooth edges. As \cite{T16} argued, if dust has a radial variation in the size distribution, assuming uniform dust properties across the disk can result in a shallower profile. This has been confirmed by \cite{MC12}, who showed that GQ Lup A's disk does have a radial gradient in both the dust composition and grain size, with larger grains in the inner disk and submicron grains in the outer disk. The smaller disk size measured at 1.3 mm compared to 870 \micron~provides further evidence that larger grains are present in the inner disk.

\subsection{GQ Lup B's Disk Mass}
The disk around GQ Lup B is undetected in our map. As shown in Figure \ref{fig:upper_limits}, our data put a upper limit on the disk mass for GQ Lup B at $<$0.25--1 $M_\earth$ from a point source to 1/3 of the Hill radius, although this is not as strong as the upper limit of $<$0.04 $M_\earth$ by \cite{M16}. Unlike another wide-orbit substellar companion FW Tau C, which has 1 to 2 $M_\oplus$ of dust in its disk \citep{K15}, GQ Lup B's disk appears to have little dust, similar to the dust-depleted disk around GSC 6214-210 B ($<$0.15 $M_\earth$ of dust; \citealt{B15}). This may arise from different evolutionary stages: FW Tau C is younger ($\sim$2 Myr) and has a more massive accretion disk, while GQ Lup B (2--5 Myr) and GSC 6214-210 B (5--10 Myr) are more evolved and their disks are rather depleted.

\begin{figure*}[t]
\centering
\includegraphics[angle=0,width=0.33\linewidth]{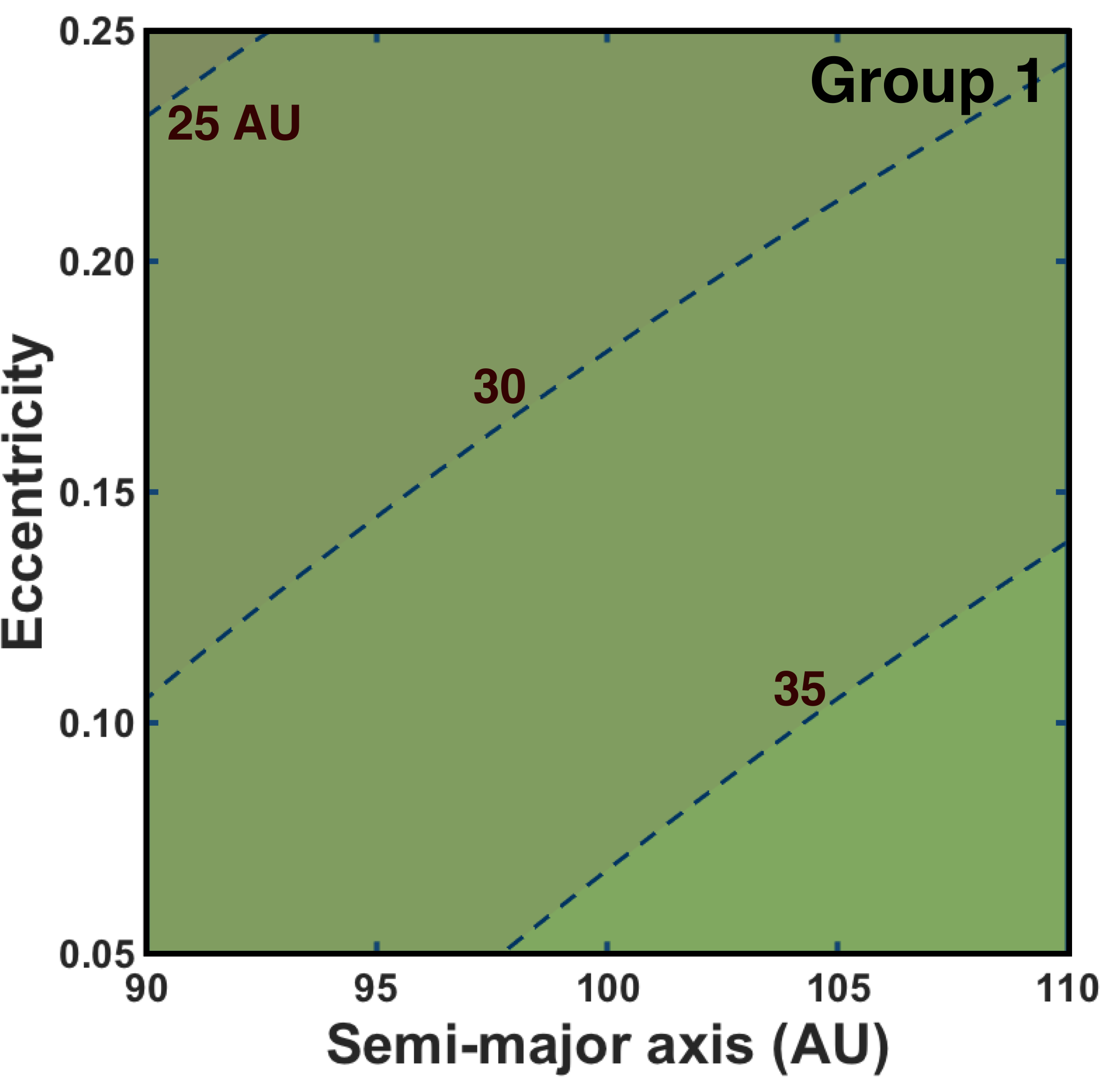}
\includegraphics[angle=0,width=0.33\linewidth]{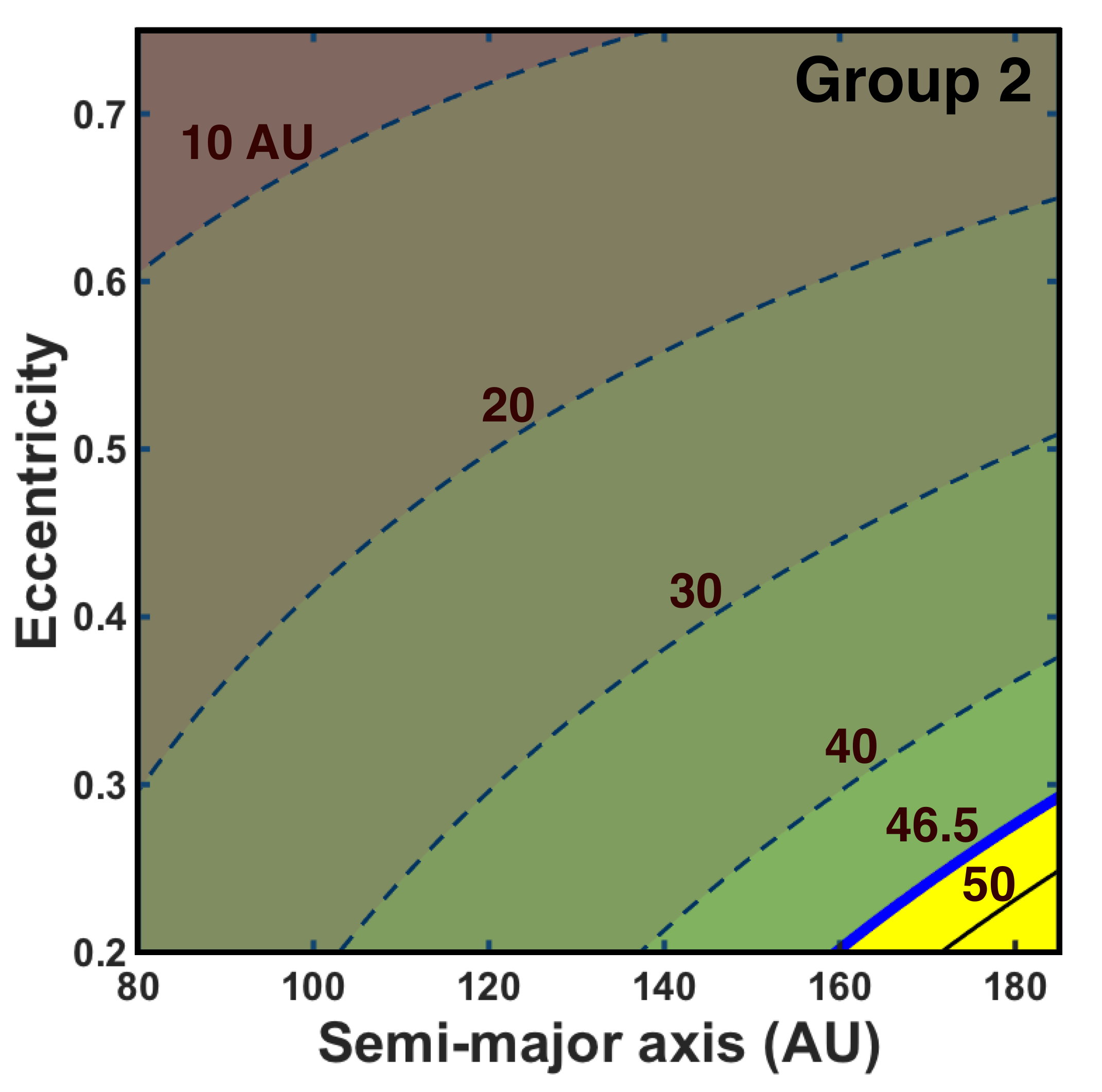}
\includegraphics[angle=0,width=0.33\linewidth]{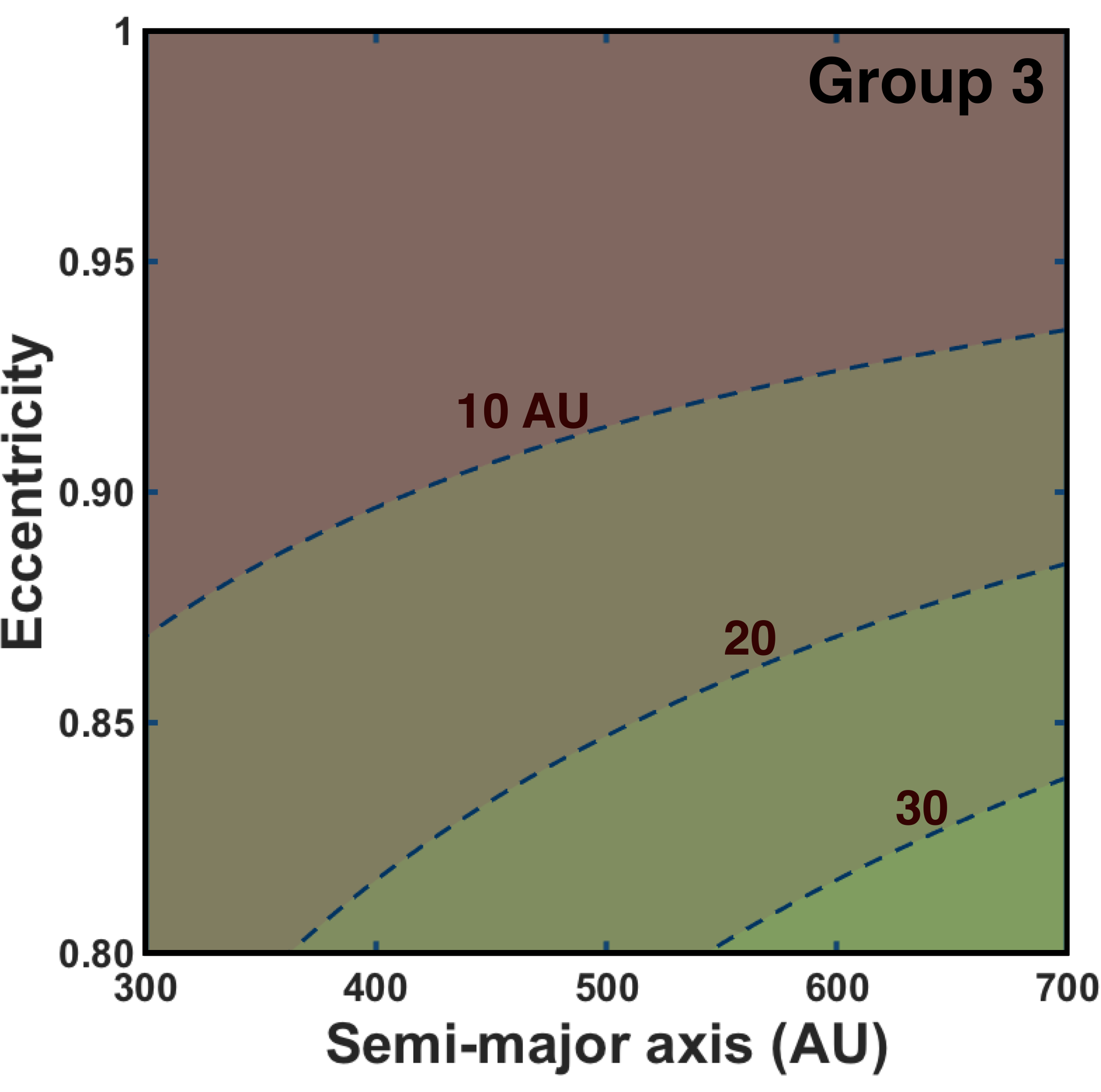}
\caption{Tidal truncation radius of GQ Lup A's disk as a function of the semi-major axis and eccentricity of GQ Lup B's orbit for the three groups of possible orbits in \cite{S16}. Contours represent lines of constant truncation radius. Orbits should have a tidal truncation radius greater than $\sim$46.5 AU to be consistent with the measured size of GQ Lup A's gas disk. Therefore, Group 1 and 3 are probably rejected since their truncation radii are less than 40 AU. GQ Lup B's orbit may have $a\sim170$ AU and $e\sim0.25$, as shown in the yellow region of Group 2.
}
\label{fig:tidaltruncation}
\end{figure*}

\begin{figure*}[t]
\centering
\includegraphics[angle=0,width=0.495\linewidth]{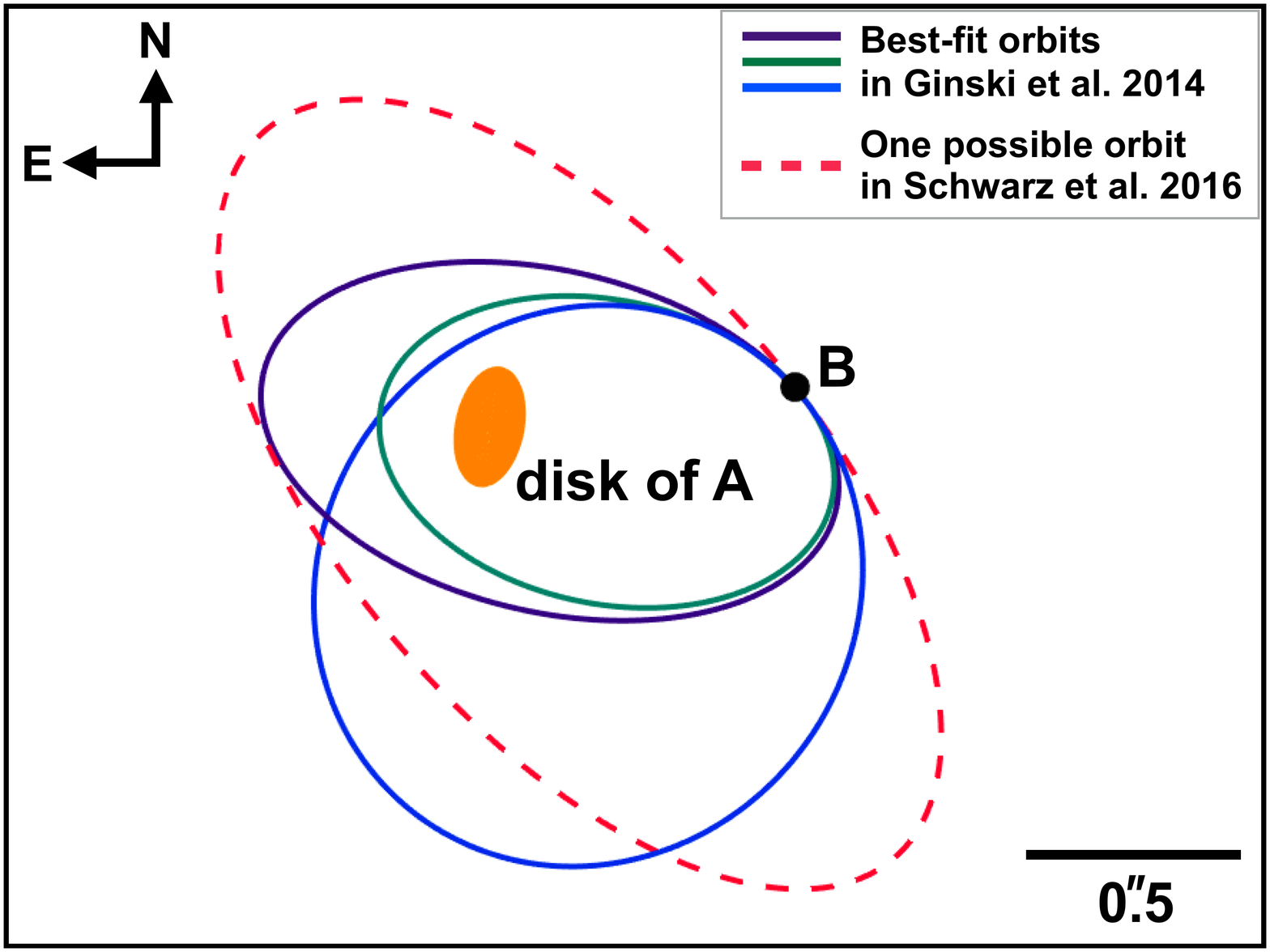} 
\includegraphics[angle=0,width=0.495\linewidth]{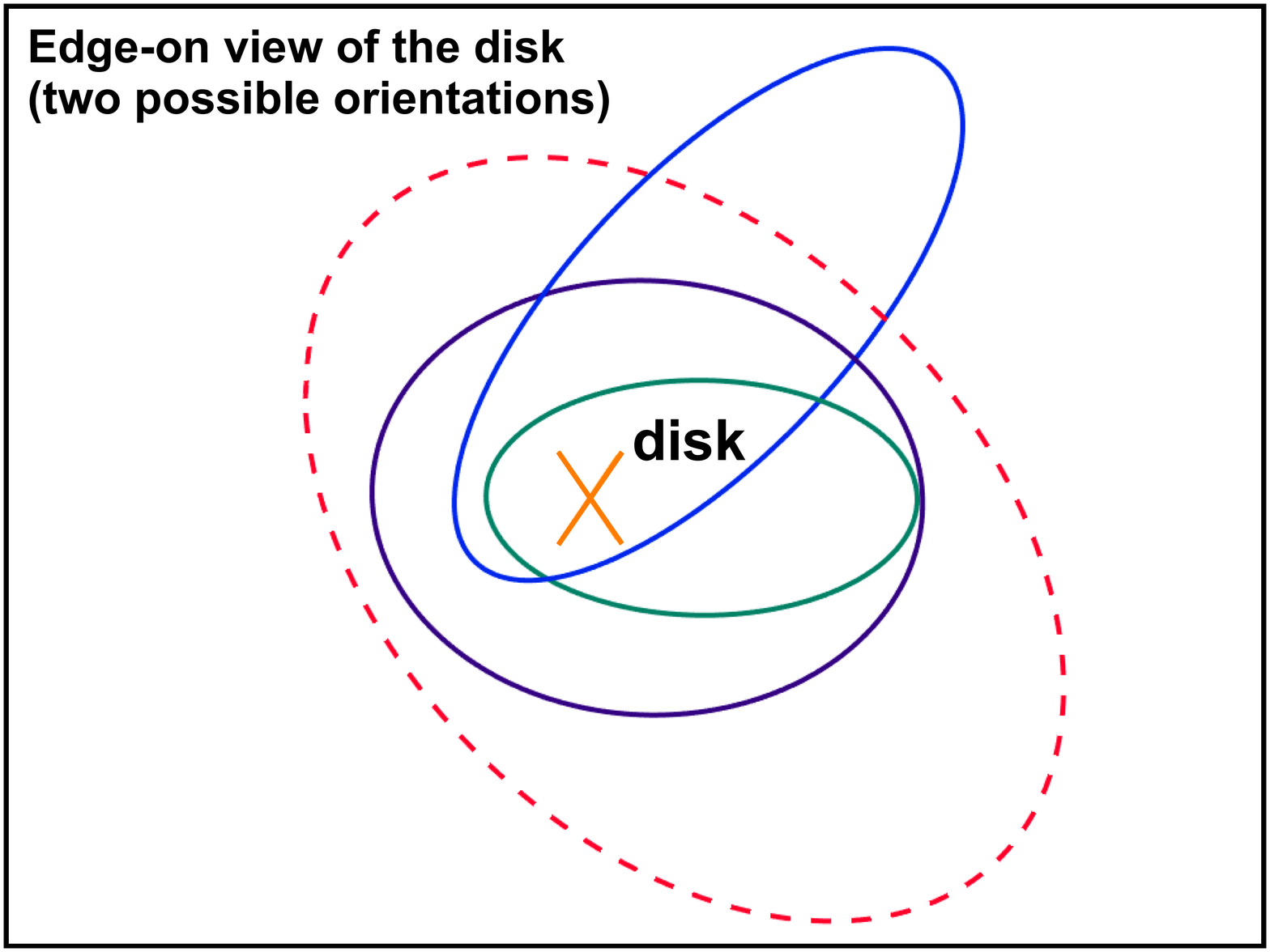}
\caption{Disk of GQ Lup A and orbital solutions of GQ Lup B. Left: when the system is projected on the sky plane. We plot the three best-fit orbits in \cite{G14}. We also plot one possible solution from \cite{S16}, with $a\sim170$ AU and $e\sim0.25$ constrained by our analysis in Figure \ref{fig:tidaltruncation}. Right: when the disk is viewed edge-on. Two possible disk orientations with identical inclination relative to the line of sight are plotted for comparison. None of these orbits is coplanar with the disk.}
\label{fig:orbits}
\end{figure*}

\begin{figure}[h]
\centering
\includegraphics[angle=0,width=\columnwidth]{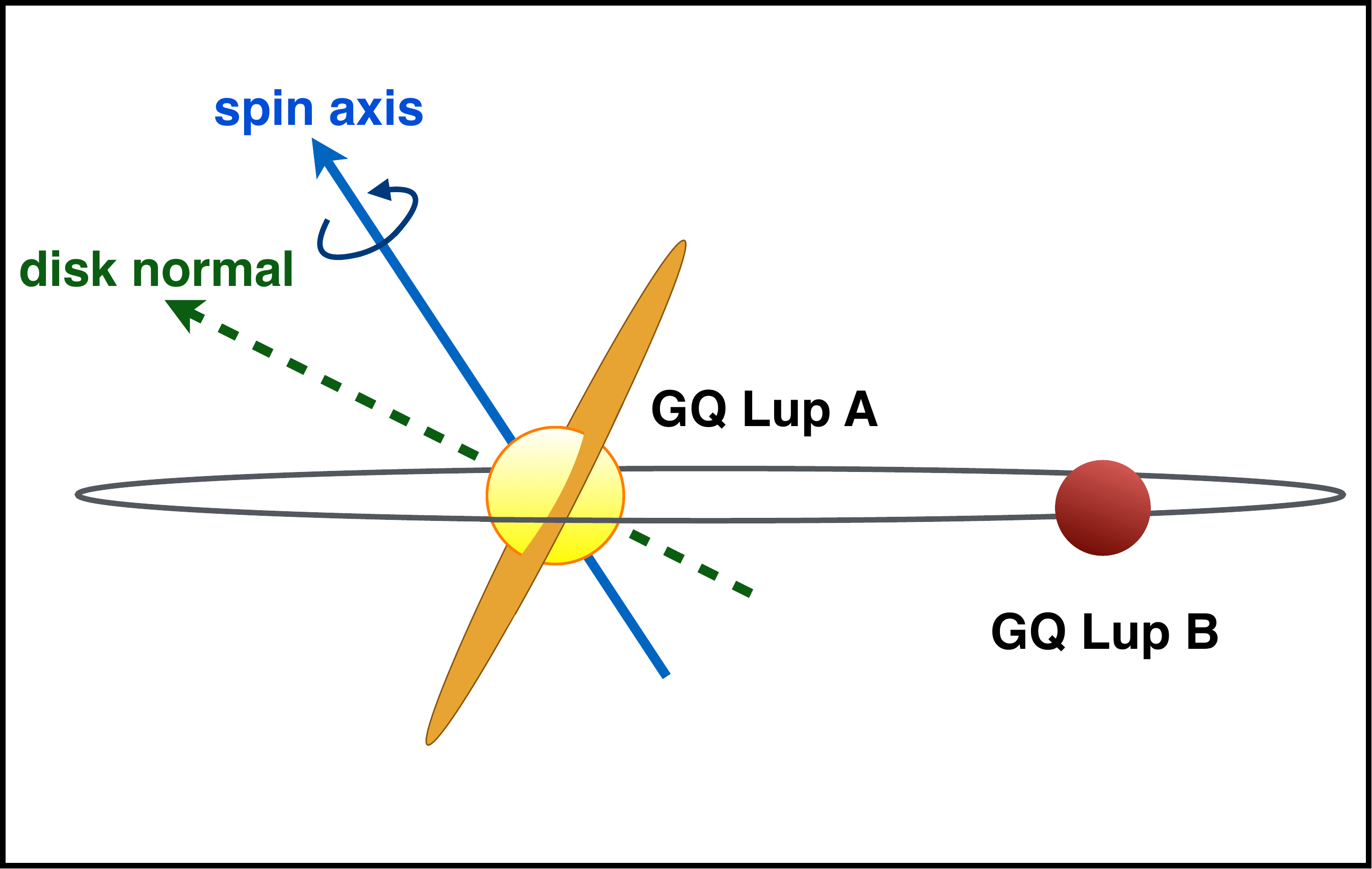}
\caption{Our rendition of the GQ Lup system. GQ Lup B is perhaps on a low eccentricity orbit not coplanar with the circumstellar disk. In addition, GQ Lup A's spin axis is not aligned with the disk ($i\sim27\degr$ versus $\sim$56\degr). Sizes are not to scale.}
\label{fig:geometry}
\end{figure}

\subsection{Orbital Constraints from Disk Size}
\label{Section:tidaltruncation}
GQ Lup B's orbital motion was first detected by \cite{G14}, who showed that the best-fit orbits were eccentric. With new RV measurements, \cite{S16} showed that the semi-major axis $a$, eccentricity $e$, and inclination $i$ of GQ Lup B's orbit fall into three groups:\\
\indent 1. $a\sim$ 100 AU, $e\sim 0.15$, $i\sim57\degr$,\\
\indent 2. $a<$ 185 AU, $0.2<e<0.75$, $28\degr<i<63\degr$, and \\
\indent 3. $a>$ 300 AU, $e>$ 0.8, $52\degr<i<63\degr$.\\
As was argued by \citet{S16}, Group 3 is a priori unlikely as its high eccentricity and long orbital period would mean that we are observing GQ Lup B close to periastron. Since the circumstellar disk may be disrupted if the companion goes too close to the star, we can calculate the truncation radius for each of the orbit groups to determine whether they are consistent with the size of the GQ Lup A disk. We adopt a disk radius of 46.5 AU determined by \cite{M16} from CO (3--2) emission, as it more likely represents the full extent of the disk compared with our 1.3 mm size measurement.

Semi-analytic approximations for the tidal truncation radius of the primary star as a function of orbital parameters find that the disk should be truncated at a radius of
\begin{equation}
R_t \approx 0.36 \left[\frac{(1-e)^{1.2} \, \phi^{2/3} \, \mu^{0.07}}{0.6\, \phi^{2/3} + \ln{(1 + \phi^{1/3})}}\right] \, a,
\end{equation}
where $\phi$ is the ratio of the mass of the primary to the mass of the secondary, and $\mu \equiv 1/(1+\phi)$ \citep{E83,P05}. This equation is relatively insensitive to whether the mass of GQ Lup B is closer to 10 $M_{\text{Jup}}$ or 40 $M_{\text{Jup}}$ (because it is at most 4\% of the primary's mass), but is very sensitive to eccentricity and semi-major axis. We hence arbitrarily use 25 $M_{\text{Jup}}$ for this calculation. 

We show the truncation radius as a function of semi-major axis and eccentricity in Figure \ref{fig:tidaltruncation}. We find that the truncation radii for Group 1 and Group 3 are all less than 40 AU, not compatible with the measured disk size. Most of the parameter space for Group 2 is also excluded; the remaining solutions are those with $a\sim$ 160--180 AU and $e\sim$ 0.2--0.3. As a result, the size of the GQ Lup A disk suggests that GQ Lup B's orbit probably has a low eccentricity. However, it is important to note that this equation may break down if the primary's disk and secondary's orbit are substantially inclined (as may be the case for GQ Lup, see Section \ref{Section:geometry}). In this scenario, it is likely that larger disk radii than we conclude here would be acceptable. Indeed, the strong constrains we place here based on this analysis may instead be an indication of a high degree of inclination for the orbit.

\subsection{Geometry of the System}
\label{Section:geometry}
Since the inclination of GQ Lup A's disk, $\sim$56\degr, is consistent with orbital solutions in \cite{G14} and \cite{S16}, here we investigate if the star's disk and the companion's orbit can be possibly in the same plane.

In Figure \ref{fig:orbits}, we plot the three best-fit orbits in \cite{G14} (see their Table 4 for orbital parameters) as well as the GQ Lup A disk in two viewing angles, one along the line of sight and the other with the disk viewed edge-on. We also show one representative orbit from \cite{S16}, for which the semi-major axis and eccentricity are constrained by our tidal truncation analysis in Section \ref{Section:tidaltruncation}, and other parameters including the longitude of the ascending node and the argument of periastron are extracted from unpublished astrometric fitting (Ginski, C. 2016, private communication). All these orbits are unlikely in the same plane of the disk. Although A's disk and B's orbit may share similar inclinations, they probably have very different orientations in space.

In Figure \ref{fig:geometry} we plot a possible geometry of the GQ Lup system. The circumstellar disk is not aligned with the star's spin axis either, because they have different inclinations: $\sim$56\degr~for the disk and $\sim$27\degr~for the spin axis \citep{B07}. This is not unusual among T Tauri stars. \cite{AB13} showed that although stellar rotation angle is correlated with disk inclination, they are not identical but a mean difference $\sim$19\degr~in T Tauri systems. We note that this misalignment might be induced by a torque from GQ Lup B (e.g., \citealt{Batygin12,H13}).

We caution that the results presented here and in Section \ref{Section:tidaltruncation} are preliminary. As \cite{G14} and \cite{S16} stressed, many orbital solutions share similar $\chi^2$ in their orbital fitting. Future astrometric monitoring is essential to lift degeneracies and ascertain GQ Lup B's orbit.

\subsection{SED of GQ Lup B}
\label{Section:SED}
\begin{figure}
\centering
\includegraphics[angle=0,width=\columnwidth]{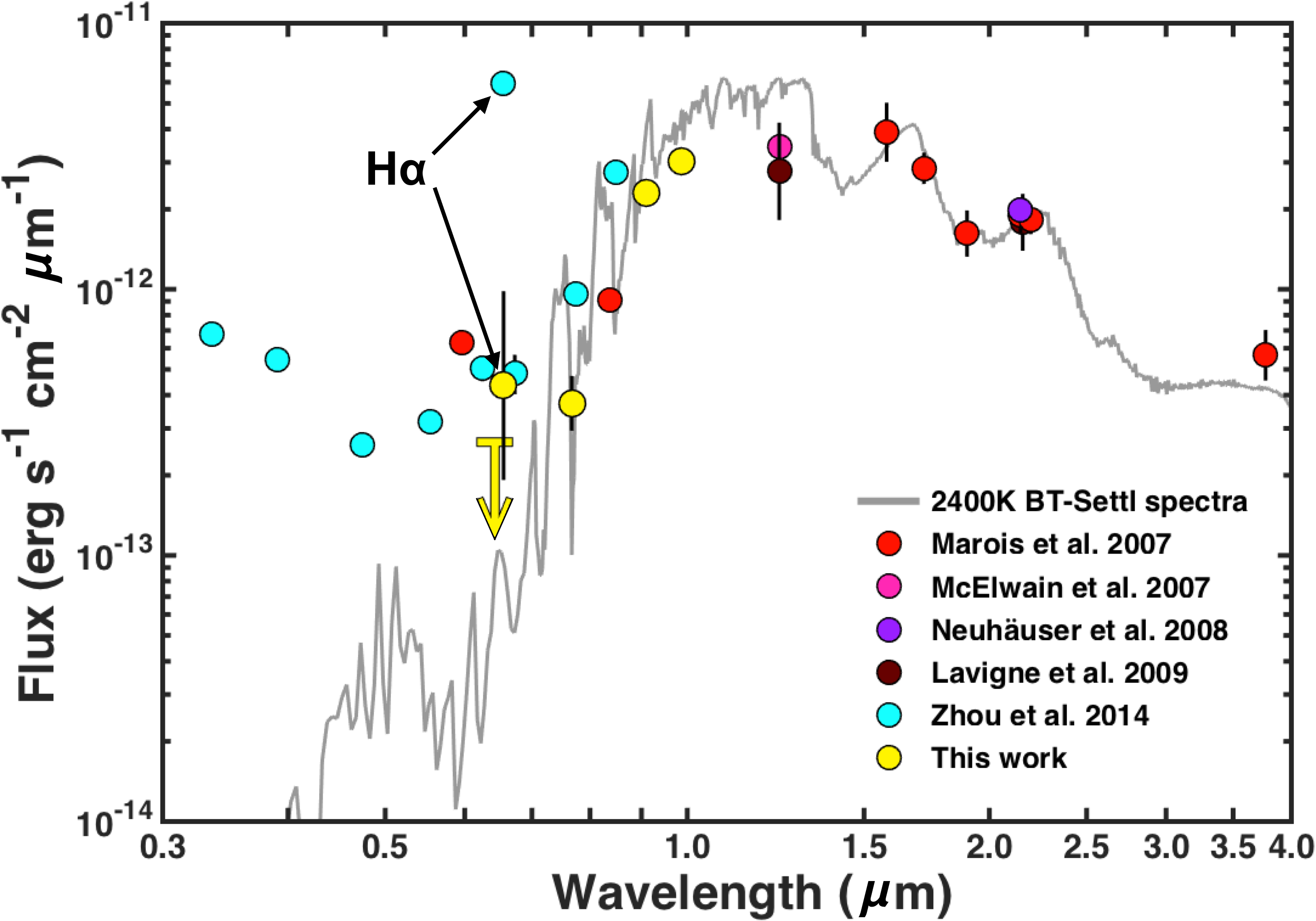}
\caption{SED of GQ Lup B. We add the observed $R$ variability amplitude of the star (0.7 mag; \citealt{B07}) to our H$\alpha$ flux uncertainty. The 3$\sigma$ limit for 0.643 \micron~also share that large uncertainty, but it is not shown for clarity. Many data have error bars smaller than the symbols.}
\label{fig:sed}
\end{figure}

Figure \ref{fig:sed} compares the spectral energy distribution (SED) of GQ Lup B to the 2400 K, log $g$ = 4.0 BT-Settl model \citep{A11}. The temperature and surface gravity are chosen to be consistent with previous estimates (e.g., \citealt{MMB07}; \citealt{M07}; \citealt{L09}). The model is normalized at $K$ to minimize the effects from dust emission and extinction. We include 0.3 to 3.7 \micron~literature photometry, the 3$\sigma$ upper limit of 0.643 \micron, and fluxes of H$\alpha$, $i'$, $z'$, and $Y_S$ in the figure.

Overall, the 2400 K model gives a reasonable fit longward of 0.7 \micron. At shorter wavelengths, the observed fluxes are much higher than photospheric due to excess continuum emission from accretion \citep{Z14}. The 0.656 \micron~H$\alpha$ emission is especially prominent. As recently simulated by \cite{SM17}, H$\alpha$ emission likely comes from the extended shock front on the surface of the circumsubstellar disk. Our measured H$\alpha$ flux, albeit with a large uncertainty, is about 10 times fainter than that of \cite{Z14}. Our 0.643 \micron~non-detection also indicates a much weaker accretion continuum. Therefore, accretion onto GQ Lup B seems to be very variable; we probably observed a more quiescent phase than did Zhou et al.

\subsection{Accretion Rate and Disk Lifetime}
Disk lifetime can be roughly estimated assuming a constant accretion rate, but it should be taken with caution as accretion can vary significantly. Our measured $\dot{M}$ for GQ Lup A, 10$^{-8}$ to 10$^{-7}$ $M_\sun$ yr$^{-1}$, is typical of T Tauri stars (e.g., \citealt{N06}). With $M_{\text{disk}}\sim$ $2\times10^{-4}$ $M_\sun$, GQ Lup A's disk may be depleted in a few thousand years. For the companion, with a disk mass upper limit of $\sim$10$^{-5}$ $M_\sun$ found by \cite{M16}, and an accretion rate of $\dot{M}\sim$ 10$^{-12}$ to $5\times10^{-10}$ $M_\sun$ yr$^{-1}$ derived by \cite{Z14} and our H$\alpha$ photometry, GQ Lup B's disk perhaps can continue for tens of thousands of years. Disk lifetime for GQ Lup B might be longer than that of the host star.

\section{Discussion}
\label{discussion}
\subsection{Formation of GQ Lup B: Scattering}
\cite{DS06} argued that scattering might be the most favorable scenario for GQ Lup B where it originally formed close to the star, but was scattered outward by a more massive body. Search for a close-in massive companion to the star have not yielded positive results. Deep AO imaging in \cite{N08} excluded any object as bright as GQ Lup B outside $\sim$18 AU. Similarly, RV monitoring rejected objects massive than 0.1 $M_\sun$ within 2.6 AU \citep{B07}, but \cite{D12} speculated that a massive brown dwarf might reside only a few AU's from the star in order to explain their observed 0.4 km s$^{-1}$ RV change in two years. However, if such an inner object exists, morphology of GQ Lup A's disk can be used to infer its presence since it may sculpt a gap or hole. The inner edge of a circumbinary disk can be truncated at approximately 2 to 3 times the binary separation \citep{AL94}, corresponding to a $\sim$10 AU hole for the brown dwarf companion posited by \cite{D12}. Nevertheless, neither the 0.3 \micron~to 1.3 mm disk SED \citep{MC12} nor our ALMA 1.3 mm disk-resolved map find evidence of a gap or central clearing in the disk. Thus, it is quite unlikely that another massive body is very close to the primary star to serve as the scatterer.

In addition, scattering often induces a very eccentric orbit (e.g., \citealt{NI11}), but for GQ Lup B, orbital solutions with low eccentricities are more probable, as shown in Section \ref{Section:tidaltruncation}. Therefore, all lines of observational evidence suggest that in situ formation via disk fragmentation or prestellar core collapse is more likely the formation pathway of GQ Lup B. This is in line with the null result of the dedicated AO search for scatterers in other systems \citep{Br16}; thus, core accretion plus subsequent scattering is perhaps not responsible for most substellar companions on wide orbits.

\subsection{Formation of GQ Lup B: in situ}
Observationally distinguishing prestellar core collapse from disk fragmentation requires high-resolution imaging toward the earliest phase in star formation. Recent studies suggest that prestellar core collapse can be effective to form very wide ($>$1000 AU) binary or multiple star systems (e.g., \citealt{P15}), while disk fragmentation may form more compact systems with separations of tens to hundreds of AU between the components (e.g., \citealt{Tobin16}). At $\sim$110 AU from the host star, GQ Lup B seems to fit nicely to the disk fragmentation scenario. 

For fragmentation to occur, GQ Lup A's disk must have been very massive and presumably more extended than 50 to 100 AU, since circumstellar disks are expected to become gravitationally unstable beyond that radius (e.g., \citealt{C09}). It is sometimes suggested that the disk plane and the companion's orbital plane should be coplanar because the companion formed in the disk. However, dynamical interactions with other fragments in the parent disk can gradually alter the initial configuration, thereby creating inclined systems \citep{SW09}. As a result, even though in Section \ref{Section:geometry} we have shown that GQ Lup B's orbital plane is probably not coplanar with GQ Lup A's disk, disk fragmentation remains a possibility. 

Recently, \cite{SH15} proposed that properties of circumsubstellar disks provide an observational diagnostic to distinguish disk fragmentation from prestellar core collapse. They predicted higher disk masses and accretion rates for objects formed via disk fragmentation, because they have longer time to accrete and thus retain a more massive disk. The discrepancy between the two scenarios is more profound for very low-mass companions, especially $<$10 $M_\mathrm{Jup}$. The authors also predicted that, under the disk fragmentation framework, low-viscosity circumsubstellar disks tend to have masses and accretion rates higher than that of high-viscosity ones, because higher viscosity facilitates angular momentum transport and disk dissipation.

\begin{figure}
\centering
\includegraphics[angle=0,width=\columnwidth]{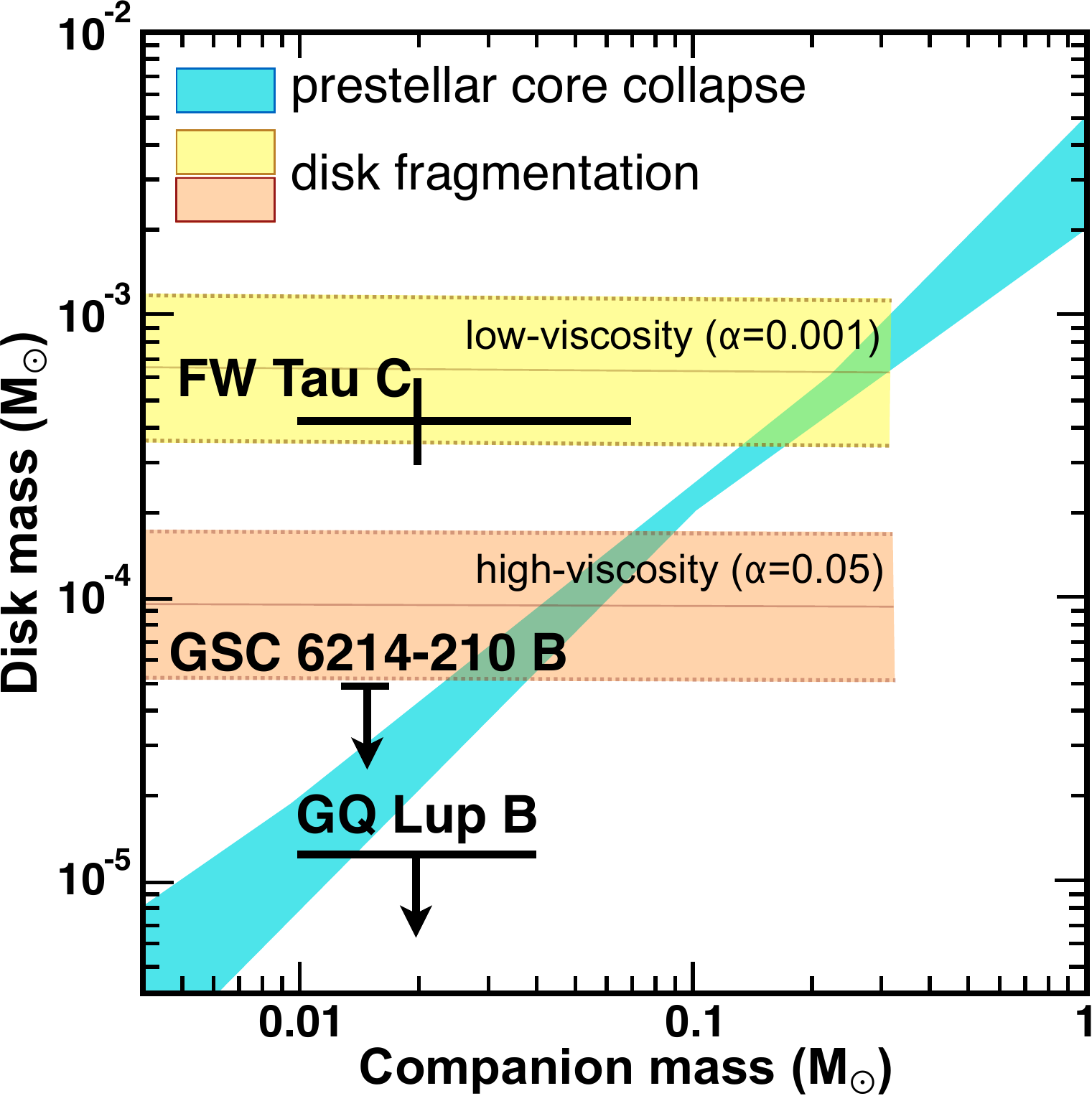}
\caption{Mass of disk versus mass of companion for two formation pathways. Figure adapted from \cite{SH15}. Turquoise area represents the best fits to the Taurus star-forming region in \cite{A13}, without including the 0.7 dex of standard deviation. Yellow and salmon swaths are the $\pm$1$\sigma$ intervals for 1 to 10 Myr objects formed by disk fragmentation. Disk masses for substellar companions are adopted from \cite{K15}, \cite{B15}, and \cite{M16}. GQ Lup B seems to be more consistent with the formation via prestellar core collapse.
}
\label{fig:GI}
\end{figure}

In Figure \ref{fig:GI} we overplot substellar companions FW Tau C, GSC 6214-210 B, and GQ Lup B on the Figure 7 of \cite{SH15}. Objects formed by disk fragmentation have a roughly constant disk mass, in drastic contrast to the monotonic correlation $M_\mathrm{disk}\propto M_\mathrm{star}$ for prestellar core collapse. While FW Tau C has a rather massive disk, considering the large dispersion in the $M_\mathrm{disk}\propto M_\mathrm{star}$ correlation ($\pm$0.7 dex; \citealt{A13}), it is still consistent with both formation scenarios. The very low-mass disks around GQ Lup B and GSC 6214-210 B are more in line with the formation via prestellar core collapse. Alternatively, if they formed via disk fragmentation, their disks might have a high viscosity to quickly dissipate.

\subsection{Formation of Satellites}
Disks around wide-orbit substellar companions provide clues to the formation and population of exomoons, which are challenging to detect with current facilities. Simulations of \cite{PL07} found that Jupiter-mass satellites are unlikely to form around brown dwarfs because no rocky cores grow fast enough to accrete a gaseous envelope before the disk dissipates, while Earth-like satellites can be common if disk mass is a few $M_\mathrm{Jup}$. Nonetheless, even Earth-mass satellites are rare if disk mass is only a fraction of $M_\mathrm{Jup}$. Thus, it appears that GQ Lup B has no gaseous moons, while a few Earth-like moons may have formed in early times when the disk was more massive. As \cite{M16} and our ALMA observations find that GQ Lup B's disk is deficient in dust, forming Earth-like satellites is no longer possible. Only tiny rocky moons analogous to the Moon ($\sim$0.012 $M_\earth$) may still form out of the remaining material ($<$0.04 $M_\earth$). Satellite formation around GQ Lup B is probably in the late stage and may have already ceased.

\section{Conclusions}
We observe the 2--5 Myr GQ Lup system with ALMA at 1.3 mm and MagAO at 0.6 to 1~\micron. With an unprecedented $0\farcs054\times0\farcs031$ resolution at 1.3 mm, we resolve GQ Lup A's accretion disk. Our observations, however, are not deep enough to detect GQ Lup B's disk. The main results are as follows.
\begin{itemize}[leftmargin=0.33cm]
\item GQ Lup A's disk has a radius of $\sim$22 AU, a dust mass of $\sim$6 $M_\earth$, an inclination angle of $\sim$56\degr, a position angle of $\sim$349\degr, and a very flat surface density profile. The flat profile is indicative of radial variation of dust sizes, with larger grains growing in the inner disk. This is also supported by the larger disk size measured at a shorter wavelength of 870 \micron~\citep{M16}.
\item GQ Lup A's disk is not aligned with the star's spin axis ($i\sim56\degr$ versus $\sim$27\degr), and it is unlikely to be coplanar with GQ Lup B's orbit. We use the size of the GQ Lup A disk to demonstrate that GQ Lup B's orbit might have a low eccentricity $e\sim$ 0.2--0.3 with semi-major axis $a\sim$ 160--180 AU. Highly eccentric orbits have tidal truncation radii incompatible with the measured disk size.
\item Both components are glowing in H$\alpha$, indicating active accretion. We derive accretion rates of $\dot{M}\sim10^{-8}$ to $10^{-7}~M_\sun~\mathrm{yr}^{-1}$ for GQ Lup A, and $\dot{M}\sim10^{-12}$ to $10^{-11}~M_\sun~\mathrm{yr}^{-1}$ for GQ Lup B. This implies that GQ Lup A's disk may be depleted in a few thousand years, while GQ Lup B's disk may remain longer.
\item Both our disk modeling and the more sensitive observations by \cite{M16} suggest that GQ Lup B's disk is rather dust-depleted, similar to GSC 6214-210 B ($<$0.15 $M_\earth$ of dust; \citealt{B15}), but in contrast to the dust-abundant disk around FW Tau C (1--2 $M_\earth$ of dust; \citealt{K15}). This may be due to age differences, as GQ Lup and GSC 6214-210 are old compared with FW Tau.
\item Since there are no gaps or an inner cavity in GQ Lup A's disk, the chance of having another inner companion more massive than GQ Lup B is low. Therefore, scattering is unlikely responsible for GQ Lup B's formation; in situ formation via disk fragmentation or prestellar core collapse is favored. The very low-mass disk of GQ Lup B is more consistent with prestellar core collapse based on the simulations in \cite{SH15}.
\item Based on the results of \cite{PL07}, GQ Lup B probably has no gaseous satellites. With very little dust remaining in the disk, only tiny rocky moons might form around GQ Lup B.   
\end{itemize}

\acknowledgements
We thank the referee for helpful comments. We are grateful to Christian Ginski and Henriette Schwarz for providing their new astrometric fitting, and Yifan Zhou for the H$\alpha$ contrast in the $HST$ data. We thank Kaitlin Kratter, Min-Kai Lin, Yu-Cian Hong, and Jing-Hua Lin for discussions. We are also grateful to the MagAO development team and the Magellan Observatory staff for their support. This material is based upon work supported by the National Science Foundation under Grant No. 1506818 (PI Males) and NSF AAG Grant No. 1615408 (PI Close). Y.-L.W. and L.M.C. are supported by the NASA Origins of Solar Systems award and the TRIF fellowship. J.R.M. and K.M.M. were supported under contract with the California Institute of Technology (Caltech) funded by NASA through the Sagan Fellowship Program. K.M.M's and L.M.C's work is supported by the NASA Exoplanets Research Program (XRP) by cooperative agreement NNX16AD44G. This paper makes use of the following ALMA data: ADS/JAO.ALMA\#2015.1.00773.S. ALMA is a partnership of ESO (representing its member states), NSF (USA) and NINS (Japan), together with NRC (Canada), NSC and ASIAA (Taiwan), and KASI (Republic of Korea), in cooperation with the Republic of Chile. The Joint ALMA Observatory is operated by ESO, AUI/NRAO and NAOJ. The National Radio Astronomy Observatory is a facility of the National Science Foundation operated under cooperative agreement by Associated Universities, Inc. Results from distributed computing were obtained using the Chameleon testbed supported by the National Science Foundation.

\end{document}